\documentstyle[12pt,aasms4]{article}

\input epsf.sty
\epsfverbosetrue

\def\drvf#1#2{{{d{#1}}\over{d{#2}}}}

\def\gtaprx{\mathrel{\vcenter{\offinterlineskip \hbox{$>$}}}}

\newcommand{\sig}{\:\lower0.6ex\hbox{$\stackrel{\textstyle >}{\sim}$}\:}
\newcommand{\sil}{\:\lower0.6ex\hbox{$\stackrel{\textstyle
<}{\sim}$}\:}
\newcommand{\sigs}{\:\lower0.4ex\hbox{$\stackrel{\scriptstyle
       >}{\scriptstyle \sim}$}\,}
\newcommand{\sils}{\:\lower0.4ex\hbox{$\stackrel{\scriptstyle
       <}{\scriptstyle \sim}$}\,}

\begin{document}

\title{Nucleosynthesis in Neutrino-Driven Winds: II. Implications
for Heavy Element Synthesis}

\author{R. D. Hoffman and S. E. Woosley}
\affil{Board of Studies in Astronomy and Astrophysics, University of
California, Santa Cruz, CA 95064}

\and

\author{Y.-Z. Qian}
\affil{Physics Department, 161-33, California Institute of Technology,
Pasadena, CA 91125}

\begin{abstract}

During the first 20 seconds of its life, the enormous neutrino luminosity
of a neutron star drives appreciable mass loss from its surface. This 
neutrino-driven wind has been previously identified as a likely site for the
$r$-process. Qian \& Woosley (1996) have derived, both
analytically and numerically, the physical conditions relevant for heavy 
element synthesis in the wind. These conditions include
the entropy ($S$), the electron fraction ($Y_e$), the dynamic time scale,
and the mass loss rate.
Here we explore the implications of these conditions
for nucleosynthesis. We find that the standard wind models derived in that
paper are inadequate to make the $r$-process, though they do produce some rare
species above the iron group. We further determine the general
restrictions on the entropy, the electron fraction, and the dynamic time scale
that {\sl are} required
to make the $r$-process. In particular, we derive 
from nuclear reaction network calculations the   
conditions required to give a sufficient neutron-to-seed ratio
for production of the platinum peak. 
These conditions range from 
$Y_e\approx 0.2$ and  
$S\sil 100$ per baryon for reasonable
dynamic time scales of $\sim 0.001$--0.1 s,
to $Y_e\approx 0.4$--0.495 and  
$S\sig400$ per baryon for a dynamic time scale of $\sim 0.1$ s. 
These conditions are also derived analytically to illustrate the physics
determining the neutron-to-seed ratio.   

\end{abstract}


\section{Introduction}

The origin in nature of the neutron-rich isotopes 
heavier than the iron group remains uncertain,
but rapid progress in both theory and observation has been
made in recent years. For example, 
analysis of the metal deficient star CS 22892-052 ([Fe/H] $\approx
-3.1$) by Sneden et al. (1996) shows
striking evidence that the solar abundance
pattern for the $r$-process isotopes exists even at very early times
in our Galaxy, and suggests an origin for these nuclei
quite distinct from the $s$-process. The fact that the $r$-process
abundance pattern in this very metal deficient star is so strikingly solar
across the entire range $56 \le Z \le 76$ also suggests that this
pattern is generic and reflects the conditions constantly obtained
in a unique astrophysical environment responsible for the $r$-process.
However, we note that the current observational data for $Z < 56$ 
are not so conclusive.

On the theoretical front, many recent calculations (e.g., \cite{wh92};
\cite{how93}; \cite{twj94}; 
\cite{wwmhm94}) have shown that a promising site for $r$-process 
nucleosynthesis
is the neutrino-driven wind blowing from a nascent neutron star following
a core-collapse-driven supernova. This $r$-process site has some
attractive features. First of all, the $r$-process would be primary, 
in accordance
with the observation of \cite{sneden96} (see also \cite{cow96};
\cite{matcow90}; \cite{ctt91}). 
Furthermore, the total mass loss in the wind
approximately accounts for the amount of
$r$-process material ejected per supernova, $\sim 10^{-5}\ M_{\odot}$, 
as expected from the supernova rate and total $r$-process yields of the Galaxy.
Finally, the conditions in the wind are determined by
the properties
of the neutron star and the characteristics of its neutrino emission.
Therefore, the $r$-process nucleosynthesis might be approximately constant
from event to event, reflecting the near constancy of the neutron star
mass and cooling history, if possible complications due to fallback 
could be ignored. 

However, problems have emerged in the theoretical
model. The successful $r$-process calculations, with the possible exception of
Woosley et al. (1994),
all utilize parametric modifications to the key parameters,
especially an artificial increase in
the entropy. The high entropies of Wilson's supernova model reported
in Woosley et al. (1994) have not been replicated elsewhere.
This discrepancy has been
emphasized by the analytic calculations of Qian \& Woosley
(1996, hereafter Paper I). In general, the deficiency in
entropy is only a factor of two, but the gap is proving difficult to
bridge. Paper I also suggested that other relevant parameters of the
problem, especially a short dynamic time scale or 
large neutron excess, might
compensate for the low entropy, or that there may emerge other sources
of entropy hitherto neglected. But, for the time being, a potentially
beautiful solution to a classic problem --- the origin of 
the $r$-process --- falters by a factor of two.

We will not resolve this quandary in the present paper. What
we shall do however, is: a) show that the standard models of
the neutrino-driven wind 
derived in Paper I give interesting nucleosynthesis above
the iron group, but with the electron fractions obtained in 
Woosley et al. (1994), do not give the classical $r$-process; and b) determine, both
analytically and numerically, the conditions that {\sl are} required
in this sort of model to produce the
$r$-process, in particular the platinum peak.

\section{Nucleosynthesis In Neutrino-Driven Winds}

An $r$-process might occur for various combinations of
entropy, electron fraction, and dynamic time scale.
Different authors, using numerical supernova models, 
have arrived at qualitatively different values for 
these key parameters in the neutrino-driven wind. 
In order to obtain a better understanding
of the physical conditions,
both analytic and numerical studies of the wind were carried out in Paper I. 
The primary goal of that paper was to
examine the dependencies of the physical parameters in the wind
on the neutron star mass ($M$) and radius ($R$), and the 
emergent neutrino luminosity ($L_{\nu}$) and energy spectra. 
The analytic study was in the same spirit as that of 
Duncan, Shapiro, \& Wasserman (1986),
but was more extensive and directed towards nucleosynthetic issues. 
The analytic results of Paper I were given by equations
(48a), (48b), and (49) for the entropy ($S$), 
equations (58a) and (58b) for the mass outflow rate ($\dot M$), 
equation (61) for the dynamic
time scale ($\tau_{\rm dyn}$), and equation (77) for 
the electron fraction ($Y_e$).

Physically, the first three parameters ($S$, $\dot M$, and $\tau_{\rm dyn}$)
are determined by the sum of heating produced by all neutrinos,
whereas the evolution of $Y_e$ mostly reflects the difference between the
$\nu_e$ and $\bar\nu_e$ fluxes through the inter-conversion of
free nucleons by $\nu_e$ and $\bar\nu_e$ captures. 
As discussed in Paper I, the first three
parameters are not sensitive to the exact values of $Y_e$, and 
their determination can be essentially
decoupled from the evolution of $Y_e$ in the wind.
The analytic results for these three parameters were tested by
a series of numerical calculations
using the one-dimensional implicit hydrodynamic code KEPLER.
A total of nine models were studied. The neutron star mass and radius,
and neutrino luminosity at the inner boundary of 
these models were varied to observe the corresponding 
effects on these three parameters.
A comparison between analytic and numerical results was summarized 
in Table 1 of Paper I. 

From these numerical models, we have extracted the velocity ($v$),
density ($\rho$), and temperature ($T$)	of a mass element as functions 
of its position ($r$) in a steady-state wind. Using $dt=dr/v(r)$,
we can obtain the evolution of $r$, $\rho$, and $T$ with time $t$ for
the mass element. Starting at $T_9\approx10$ ($T_9$ is the
temperature in units of $10^9$ K), when nuclear
statistical equilibrium (NSE) is assured, we then follow the time
evolution of the nuclear composition in this mass element with 
a reaction network (\cite{wh92}), until
the composition freezes out at $T_9\approx 1$.
To good accuracy, the mass element is initially composed of free nucleons, 
in proportions specified by the initial electron fraction
$Y_{e,i}$. The initial electron fraction is essentially determined by
the equilibrium between $\nu_e$ and $\bar\nu_e$ captures on free
neutrons and protons, respectively (\cite{qian93}; see also Paper I).
If we denote the rates for $\nu_e$ and $\bar\nu_e$ captures on 
free nucleons as $\lambda_{\nu_en}$ and $\lambda_{\bar\nu_ep}$, 
respectively, the initial electron fraction is given by
\begin{equation}
Y_{e,i} = { 1\over{1+\lambda_{\bar\nu_e p}/\lambda_{\nu_e n} }}.
\end{equation}
In turn, the rate $\lambda_{\nu_e n}$ ($\lambda_{\bar\nu_e p}$) is determined
by the $\nu_e$ ($\bar\nu_e$) luminosity and energy spectrum.

In realistic supernova models, the neutrino luminosity and energy spectra
evolve with time. The individual wind models of Paper I represent the
steady-state configurations reached at different neutrino luminosities
over time scales much shorter than the evolution time scales of the
neutrino luminosity and energy spectra. Because the time evolution of
the neutrino energy spectra is much less pronounced than that of the
neutrino luminosity, two generic sets of neutrino mean energies were
assumed for these models. From the analytic results of Paper I, we
can see that the mass loss rate, the dynamic time scale, and especially
the entropy would not change significantly if a more precise prescription
of the neutrino energy spectra were used. However, because the electron
fraction is sensitive to the difference between the $\nu_e$ and $\bar\nu_e$
fluxes, and the nature of heavy element nucleosynthesis is extremely
sensitive to $Y_e$, in this paper
we calculate the $\nu_e$ and $\bar\nu_e$ reaction
rates according to the time evolution of neutrino luminosity and 
energy spectra in Wilson's 20 M$_\odot$  supernova model used in
Woosley et al. (1994). Specifically, we take 
from Wilson's supernova model the neutrino energy spectra
corresponding to the same neutrino luminosity as in
the wind model. Starting with the initial value in equation (1),
we then follow the evolution of $Y_e$ in the reaction network, taking into
account $\nu_e$ and $\bar\nu_e$ captures on free nucleons and heavy nuclei
(\cite{macful95}),
electron and positron captures on free nucleons and heavy nuclei, 
and nuclear $\beta$-decays.

The results of the nucleosynthesis calculations are presented in 
Tables 1--3. 
The entropy, the initial 
electron fraction, 
and the dynamic time scale are given for each wind model.
The dynamic time scale roughly corresponds to the time over which
the temperature changes by one $e$-fold (Paper I). In order to
check the influence of the neutrino flux on the nucleosynthesis,
we have carried out five different calculations.
All runs included electron and positron captures on free nucleons
and nuclei, as well as nuclear $\beta$-decays. The individual runs
differ in the inclusion of various neutrino reactions. Respectively,
they cover the cases
including (1) no neutrino reactions (column 2), (2) $\nu_e$ and
$\bar\nu_e$ captures on free nucleons only (column 3), 
(3) $\nu_e$ and $\bar\nu_e$ captures on free nucleons, and 
neutral-current neutrino spallation on $\alpha$-particles (column 4),
(4) $\nu_e$ and $\bar\nu_e$
captures on free nucleons and nuclei (column 5), 
and (5) $\nu_e$ and $\bar\nu_e$
captures on free nucleons and nuclei, as well as neutral-current
neutrino spallation on $\alpha$-particles (column 6).
For all runs, we give the electron fraction ($Y_{e,f}$), the
average mass number of nuclei excluding free nucleons and 
$\alpha$-particles ($\bar A$), and the neutron and $\alpha$-particle
mass fractions ($X_{n,f}$ and $X_{\alpha,f}$) at the freeze-out of
the charged-particle reactions ($T_9\approx 2.5$). 

In Figures 1--9, detailed nucleosynthesis results from the runs
that included $\nu_e$ and $\bar\nu_e$ captures on free nucleons (column 3
in Tables 1-3) are given 
in terms of the production factor, defined as the final mass
fraction (after all weak decays) of a given stable nucleus divided by its
solar abundance (\cite{ag89}). These results are also representative of the 
nucleosynthesis obtained in the other runs. 
In these figures, the most abundant
isotope in the solar abundance distribution for a given element is plotted as
an asterisk. Isotopes of a given element are connected by solid lines. A
diamond around a data point indicates 
that the isotope is produced chiefly as a
(neutron-rich) radioactive progenitor. The dotted horizontal
lines represent an approximate
``normalization band,'' bounded from above by the largest
production factor in the calculation and from below by a production 
factor four times smaller. Nuclei
that fall within this band will be the dominant species produced.
In Figures 1--9, no re-normalization has been attempted. 

Models 10A--F produced interesting nucleosynthesis representative
of the $\alpha$-process (\cite{wh92}; \cite{wjt94}).
The most abundant nuclei produced
have mass numbers $90 \leq A \leq 110$.
Model 10A ($Y_e\approx 0.47$, $S\sim 70$ per baryon) shows the 
production of the $N=50$ 
closed-neutron-shell nuclei which were grossly overproduced in
previous studies. Models 10B and C, 
with progressively lower neutrino luminosity
and lower values of $Y_e$, made heavier nuclei. Production of Sn, Sb, and Te
was not accurately calculated, as the radioactive progenitors for these
species were isotopes of Ru, the last element in our reaction network. 
Models 30A--C       
produced nuclei near the iron group. Model 30A exhibits interesting
nucleosynthesis for $Y_e > 0.5$, while Model 30C shows the
production of $^{64}$Zn (made as itself), the dominant isotope of
this element. This nucleus was not accounted for in the surveys of Galactic
chemical evolution and nucleosynthesis in massive stars
(\cite{tww95}; \cite{wowev95}), 
and appears to be made predominantly under conditions similar to those
obtained in the neutrino-driven wind (\cite{hwfm96}).

From Tables 1--3, 
it is also clear that inclusion of the neutrino 
reactions made a difference. The $\nu_e$ and $\bar\nu_e$ captures on
free nucleons have the largest effect. The electron fraction
increases appreciably due to these capture reactions when free nucleons
are being assembled into $\alpha$-particles.
This so-called ``$\alpha$-effect'' (\cite{fulmey95}; \cite{mfw96})
is evident when we compare the case including no neutrino reactions (column 2)
with the cases including various neutrino reactions (columns 3--6).
The inclusion of neutral-current neutrino spallation on $\alpha$-particles 
and $\nu_e$ and $\bar\nu_e$ captures on heavy nuclei did
not have a major effect on the nucleosynthesis, 
at least before the freeze-out of the charged-particle reactions.
For the relatively low entropies studied here,
the neutrino spallation on $\alpha$-particles did not have 
any appreciable influence on the final $\alpha$-particle mass fraction. 
This is to be contrasted with the dramatic effect of these spallation
reactions on the $r$-process in Wilson's high-entropy supernova model
(\cite{mey95}). The $\nu_e$ captures on heavy nuclei
may have important consequences for the ensuing 
neutron-capture phase of the $r$-process (\cite{nadpan93}).
Regardless of these issues, it is clear that none of these wind
models produced an $r$-process. The neutrons had been consumed
by the time the charged-particle reactions froze out 
($T_9\approx 2.5$),
and the nuclear flow did not even reach the $A\sim 130$ $r$-process peak.
 
Another important issue concerns the overall ejection
of the synthesized material in the wind into the interstellar medium. 
As the following simple argument will show,
the magnitudes of the production factors may preclude the
ejection of such material in all nine wind models. 
From the time evolution of neutrino luminosity in Wilson's
supernova model, we find that, for example, the neutrino luminosity
decreases from twice to half the value in Model 30B over a time of
$\tau_\nu\sim 1$ s. With a mass loss rate of $\dot M\approx 1.1\times
10^{-2} \ M_\odot \ {\rm s}^{-1}$ and the largest 
production factor of $(X_{w}/X_{\odot})_{\rm max}\approx 2.7\times 10^4$
in Model 30B (Figure 8), 
the corresponding normalized production
factor is $\sim(\dot M\tau_\nu/20\ M_{\odot})
(X_{w}/X_{\odot})_{\rm max}\sim 15$, if the total amount of ejecta
from the supernova is $20\ M_{\odot}$. Models 30A and C give normalized
production factors of up to $\sim 30$. 
For Models 10A--F, the normalized 
production factors for the nuclei produced in the largest amount
are typically
of order 100. However, Woosley \& Weaver (1993) find that the
normalized production factor should not be much above 10 in order
for supernovae to produce the observed solar abundance of oxygen.
Therefore, these wind models cannot represent in detail what commonly
occurs in supernovae. 

It is quite possible that appropriate 
modifications of these standard wind models
can lead to the physical conditions for acceptable nucleosynthesis.
Paper I studied the effects of an additional energy
source on the entropy
and dynamic time scale in the wind.
Model 10F of Paper I was recalculated
with an additional
energy input of
$5\times 10^{47}\ {\rm erg\ s}^{-1}$ distributed uniformly in volume between
15 and 25 km. At these radii, the mass loss rate has been more or less
determined. The additional energy input represents
a moderate perturbation to the total amount of heating provided by
neutrinos ($1.2\times 10^{48}\ {\rm erg}\ {\rm s}^{-1}$). As a result,
the entropy increased from 140 to 192 per baryon,
and the dynamic time scale decreased from 0.11 to 0.022 s,
while the mass loss rate slightly increased from 
$2.8\times 10^{-6}$ to $3.7\times 10^{-6}$
$M_{\odot}$ s$^{-1}$. 

The effect on the nucleosynthesis was dramatic as shown in Table 
4. The neutron-to-seed ratio (cf. eq. [3]), 
less than 10 in all of the
unmodified models, rose to $\sim 166$. 
With an average mass number of $\bar A\sim 90$ for the seed nuclei,
uranium could be produced if all the neutrons were to be subsequently
captured.
The $A\sim 195$ $r$-process peak probably could have been produced
with less additional energy input, and hence a 
lower entropy and a longer dynamic time scale,
than assumed in Paper I. We conclude that lower values of $Y_e$ than
those calculated by Wilson in Woosley et al. (1994),
or additional energy input
like that considered in Paper I, are necessary to produce an $r$-process
in a spherically symmetric wind model.

\section{The Requisite Conditions For The $r$-Process}

From our studies of nucleosynthesis in the neutrino-driven wind
in the previous section, we learn that the standard wind models 
of Paper I fail to make an $r$-process. On the other hand, we also
see that reasonable modifications of these models can
significantly change the physical conditions in the wind, 
and therefore give
rise to a possible $r$-process. In order to better understand the
deficiencies of the standard wind models, and furthermore, to
motivate and direct physically plausible modifications of these
models, we now survey the important physical parameters required
for a strong $r$-process. (See also \cite{twj94}; \cite{tak96};
\cite{fri96}; and \cite{mey96}). 

We consider the following generic model for the $r$-process.
Neutron-rich material initially
composed of free nucleons at high temperatures ($T_9\approx9$)
adiabatically expands and cools. After nearly all the protons
are assembled into $\alpha$-particles at $T_9\approx5$, an 
$\alpha$-process occurs to burn the $\alpha$-particles into heavy
nuclei. The $\alpha$-process stops when charged-particle reactions
freeze out at $T_9\approx2.5$. The heavy nuclei produced at the end
of the $\alpha$-process then become the seed nuclei for the subsequent
rapid neutron capture process, or the $r$-process. We do not intend
to account for the full detail of the $r$-process, such as the final
abundance distribution, which becomes meaningful only in the context
of a consistent astrophysical model. What we are most interested in
is the physical conditions favorable for the production of the most
abundant $r$-process nuclei, such as those in the platinum peak of
the solar $r$-process abundance distribution. For this purpose, we
can think of the $r$-process as the transformation of seed nuclei
into $r$-process nuclei through simple addition of the available
neutrons. In this sense, the possibility of producing $r$-process
nuclei around a certain mass number depends only on the relative
abundances of seed nuclei and neutrons at the end of the 
$\alpha$-process. 

In general, the composition resulting from the $\alpha$-process
satisfies 
\begin{equation}
X_{n,f}+X_{\alpha,f}+X_s\approx1,
\end{equation}
where $X_{n,f}$ and $X_{\alpha,f}$ are the final mass fractions of
neutrons and $\alpha$-particles, respectively, and $X_s$ is the total
mass fraction of seed nuclei. If we represent the seed nuclei with a
mean proton number $\bar Z$ and a mean mass number $\bar A$, we can
define a neutron-to-seed ratio at the end of the $\alpha$-process as
\begin{equation}
{n\over s}\approx{X_{n,f}\over X_s}\bar A.
\end{equation}
In terms of this ratio, our simplified condition for making $r$-process
nuclei with mass number $A$ becomes
\begin{equation}
\label{nesrp}
{n\over s}+\bar A\approx A.
\end{equation}
We are particularly interested in the production of the platinum peak,
and will describe the numerical calculations for $A\approx 200$ in
the following.

To follow the nucleosynthesis in the adiabatically cooling material
with a nuclear reaction network,
we need the initial composition and the time evolution of temperature
and density. The initial composition at $T_9\approx9$ can be simply 
specified by the initial electron fraction $Y_{e,i}$, with the initial
mass fractions of neutrons and protons given by 
$X_{n,i}\approx 1-Y_{e,i}$ and $X_{p,i}\approx Y_{e,i}$, respectively.
Because temperature and density are related through the constant
entropy for the adiabatically cooling material, we only need to specify
the temperature as a function of time. For simplicity, we introduce
a dynamic time scale ($\tau_{\rm dyn}$) over which the temperature
changes by one $e$-fold, i.e.,
\begin{equation}
T_9(t)\approx T_9(0)\exp(-t/\tau_{\rm dyn}).
\end{equation}
With this time evolution of the temperature, the duration of the 
adiabatic expansion from $T_9\approx 9$ to 2.5 is
\begin{equation}
\label{netaut}
t_{\rm exp}\approx 1.28\tau_{\rm dyn}.
\end{equation}
Hereafter, we will refer to $t_{\rm exp}$ as the expansion time.
We further assume that the entropy is dominated by contributions from
radiation and relativistic electron-positron pairs, and is given by
\begin{equation}
\label{neent}
S\approx 3.33{T_9^3\over\rho_5},
\end{equation}
where $\rho_5$ is the density in units of 
$10^5\ {\rm g}\ {\rm cm}^{-3}$,
and $S$ is in units of Boltzmann constant per baryon. For convenience,
we frequently refer to entropy without its unit throughout this paper.

Now it is straightforward to determine the combinations of 
initial electron fraction, entropy, and expansion time, for which an
$\alpha$-process can lead to a sufficient 
neutron-to-seed ratio for production of $r$-process nuclei
with $A\approx 200$. We choose a range of expansion times
($0.005\leq t_{\rm exp}\leq 0.25$ s). For each $t_{\rm exp}$, we survey
a broad range of initial electron fractions ($Y_{e,i}=0.20$, 0.25, 0.30,
0.35, 0.40, 0.45, 0.46, 0.47, 0.48, 0.49, and 0.495). 
With a particular set of $t_{\rm exp}$
and $Y_{e,i}$, we seek through iteration the appropriate entropy
which enables the $\alpha$-process to produce a final composition 
satisfying equation (\ref{nesrp}) for $A\approx 200$. 
In our calculations, we
take into account the effects of electron and positron captures and
nuclear $\beta$-decays on the evolution of the electron fraction.

The results are given in Table 5 and Figure 10. For each successful run, 
the composition at the end of the 
$\alpha$-process are given
in terms of the final neutron and $\alpha$-particle
mass fractions and the mean proton and mass numbers for seed nuclei.
The mean proton and mass numbers were calculated
from $\bar Z=\sum Y(Z_s,A_s)Z_s/\sum Y(Z_s,A_s)$ and $\bar A=\sum
Y(Z_s,A_s)A_s/\sum Y(Z_s,A_s)$, respectively, with $Y(Z_s,A_s)$ the
number fraction of the seed nucleus with proton number $Z_s$ and
mass number $A_s$. The final electron fraction and the neutron-to-seed
ratio are also given in Table 5. 
The combinations of initial electron fraction
and entropy in the successful runs for three specific expansion times
are shown as filled circles connected by solid lines in Figure 10.
At a given $t_{\rm exp}$, values of $Y_{e,i}$ and $S$
to the left of the solid line will not give a sufficient neutron-to-seed
ratio for production of the platinum peak. In addition, the following
features of this figure are worth mentioning:

(1) Depending on the expansion time, there exist many possible 
combinations of initial electron fraction and entropy that can
produce nuclei with $A\sim 195$. The
``high entropy'' scenario ($S\gtrsim 350$ and  
$0.495\gtrsim Y_{e,i}\gtrsim 0.40$) 
corresponds to longer expansion times ($t_{\rm exp} \gtrsim 0.1$ s). 
The results for $t_{\rm exp}=0.25$ s are consistent with the conditions 
seen at late times in Wilson's supernova model (\cite{wwmhm94}),
and those employed in the successful $r$-process calculations of
Woosley et al. (1994) and Takahashi, Witti, \& Janka (1994),
although the later required an 
artificial increase in the entropy (by a factor of five) to produce 
nuclei with $A\sim 195$.

(2) Alternatively, there is a ``low entropy'' scenario ($S\lesssim 200$ and
$Y_{e,i}\lesssim 0.4$) that requires shorter expansion times
($t_{\rm exp}\lesssim
0.025$ s). For the most extreme case shown ($t_{\rm exp}=0.005$ s), the
platinum peak can be made for any $Y_{e,i}\lesssim 0.495$ if $S\sim 150$.
Such an expansion would correspond to very high velocities which
might be more appropriate to a jet than to a quasi-steady-state wind as
considered in Paper I. However, these rapid expansions may not continue
after the $\alpha$-process in order to allow enough time for the neutron
capture phase of the $r$-process. The slowing down of the expansion
could be facilitated by a massive overlying mantle in the case of a
Type II supernova.

(3) For smaller values of $Y_{e,i}$, the required entropy 
decreases regardless of the
expansion time. This merely reflects the very neutron-rich nature of the
initial composition. The results of Paper I suggest
that values of $Y_e$ below 0.3
might be very difficult to achieve in the neutrino-driven wind. 
Material with $Y_e<0.3$ would need to be ejected without any 
significant interaction with neutrinos. 

(4) For a fixed expansion time, the required value of entropy actually
{\sl decreases} as $Y_{e,i}$ increases from $\sim 0.48$ to $\sim 0.495$.  
Lower entropy translates to higher
density (cf. eq. [7]), and would normally produce more seed nuclei.
However, in these cases of high $Y_{e,i}$, the electron fraction
has even more leverage on the seed production. As $Y_{e,i}$ increases
towards 0.5, the neutron abundance decreases to vanishingly small values
when the $\alpha$-process begins
at $T_9\approx 5$. This in turn diminishes the efficiency of burning
$\alpha$-particles through the main 
reaction path bridging the unstable mass gaps at $A=5$ and 8, i.e., 
$^4{\rm He}(\alpha n,\gamma)^9{\rm Be}(\alpha,n)^{12}{\rm C}$.
Table 5 shows that these high $Y_{e,i}$ values give extreme
$\alpha$-rich freeze-outs with final $\alpha$-particle mass fractions
approaching unity and comparable mass fractions for neutrons and
heavy seed nuclei.

In deriving the above results, we have made two major assumptions:
(1) the entropy is proportional to $T^3/\rho$ with the proportionality
constant calculated for a mixture of radiation and relativistic 
electron-positron pairs, and (2) weak interactions other than electron
and positron captures and nuclear $\beta$-decays can be neglected in
the nuclear reaction network. We will discuss how our results are affected
if we drop either assumption in turn.

Due to its logarithmic dependence on the temperature and density, the
entropy of non-relativistic particles roughly stays constant over the
temperature range in our calculations ($9\gtrsim T_9\gtrsim 2.5$).
For a total entropy of $S\gtrsim 20$, the change in density essentially
maintains a constant entropy of relativistic particles as the material
adiabatically cools. At high temperatures ($9\gtrsim T_9\gtrsim 5$),
the relativistic particles include photons and electron-positron pairs,
and the corresponding entropy is given by equation (7). However, as
the temperature cools below $T_9\sim 5$, electron-positron pairs begin
to annihilate. The situation is much like the Big Bang. Eventually,
at $T_9\sim 1$, the only relativistic particles in the material are
photons, with the corresponding entropy given by 
$S\approx 1.21T_9^3/\rho_5$. In general, we can write the 
entropy in relativistic particles as 
$S\approx C(T_9)T_9^3/\rho_5$.
Because $C(T_9)$ decreases noticeably from $T_9\approx 5$ to 2.5 when
the $\alpha$-process is taking place in our calculations, we have
overestimated the density by using $C(T_9)\approx 3.33$ throughout 
the adiabatic expansion of the material. Consequently, we have
overestimated the seed production and underestimated the neutron-to-seed
ratio at the end of the $\alpha$-process.
This is especially true for the cases of high entropies where more 
time is available to produce seed nuclei.

With an accurate expression
for the entropy, the condition in equation (\ref{nesrp})
is satisfied at slightly
lower entropies than those given in Table 5 for the same initial 
electron fraction and expansion time. In fact, we have repeated our
calculations for $Y_{e,i}=0.30$, 0.40, 0.45, 0.47, and 0.49,
using an exact adiabatic equation of state to compute the density 
corresponding to a specific temperature. This equation of state
takes into account the contributions
to the entropy from radiation, electron-positron pairs, and ions.
The results are shown as filled squares in Figure 10 for three of the
expansion times explored in the initial numerical survey.
As expected, our previous results using equation (7) overestimated
the entropy required to produce the platinum peak.
When we use the exact adiabatic equation of state, the required entropy 
is lower by $\sim 10$\% for the longest expansion time 
$t_{\rm exp}=0.25$ s, 
whereas for the shortest expansion time 
$t_{\rm exp}=0.005$ s, the results are essentially unchanged.

If the $r$-process occurs in an environment with intense neutrino flux,
perhaps we should also include various neutrino interactions in the nuclear
reaction network. In fact, we could have specified a less
``generic'' model by considering an $r$-process site similar to
the neutrino-driven wind. In this case, the adiabatically expanding
material is also moving away from the neutrino source. We can define
a constant dynamic time scale as $\tau_{\rm dyn}\approx r/v$, with
$r$ the distance from the neutrino source and $v$ the expansion velocity.
The time evolution of temperature in equation (5) follows on assuming
$T\propto r^{-1}$. We can then introduce an additional parameter, e.g.,
the initial neutrino flux $\Phi_{\nu,i}$ at $T_9\approx 9$, in our
calculations. As the material adiabatically expands, the neutrino flux
it receives decreases as $r^{-2}\propto\exp(-2t/\tau_{\rm dyn})$. In principle,
using the above prescription, we can repeat our calculations 
for a range of $\Phi_{\nu,i}$
with various neutrino
interactions included in the nuclear reaction network.

From our discussions in $\S2$, we have seen that for relatively low
entropies of $S\lesssim 200$, the major role of the neutrino flux in
determining the neutron-to-seed ratio is to increase the electron fraction
by $\nu_e$ and $\bar\nu_e$ captures on free nucleons through the 
$\alpha$-effect. In addition, Meyer (1995) has shown that for high entropies
of $S\sim 400$, neutral-current neutrino
spallations on $\alpha$-particles during the
$\alpha$-process can increase the production of seed nuclei. In both cases,
inclusion of neutrino interactions tends to reduce the neutron-to-seed
ratio at the end of the $\alpha$-process, although the effects of these
neutrino interactions are less important
for shorter expansion times. Therefore, we can interpret the entropies
in Table 5 and Figure 10 as the {\sl minimum} values
required to produce the platinum peak for given sets of initial electron
fraction and expansion time. With this interpretation, we can avoid
repeating our calculations and 
complicating our results with an additional parameter $\Phi_{\nu,i}$.
Of course, in a consistent astrophysical model for the $r$-process where
intense neutrino flux exists, the exact conditions for production of, e.g.,
the platinum peak, have to be determined with full consideration of
various neutrino interactions.

\section{Analytic Treatment Of The $\alpha$-Process}

In order to provide some physical insight into what determines
the neutron-to-seed ratio, and extend our results in the previous
section to production of $r$-process nuclei in general, we present
an analytic treatment of the $\alpha$-process in this section
based on our generic $r$-process model. If we ignore possible
neutrino interactions, the electron fraction at the
beginning of the $\alpha$-process ($T_9\approx 5$) is about the
same as the initial electron fraction $Y_{e,i}$ at $T_9\approx 9$.
At $T_9\approx 5$, the material is essentially composed of free
neutrons and $\alpha$-particles for $Y_{e,i}<0.5$, with almost
all the protons already assembled into the $\alpha$-particles.
The mass fractions of $\alpha$-particles and neutrons at the
beginning of the $\alpha$-process are then approximately given by
\begin{mathletters}
\begin{eqnarray}
X_{\alpha,0}&\approx& 2Y_{e,i},\\
X_{n,0}&\approx& 1-2Y_{e,i},
\end{eqnarray}
\end{mathletters}
respectively.

The composition at the end of the $\alpha$-process ($T_9\approx 2.5$)
satisfies
\begin{equation}
{1\over 2}X_{\alpha,f}+{\bar Z\over\bar A}X_s\approx Y_{e,f},
\end{equation}
where $Y_{e,f}$ is the final electron fraction at $T_9\approx 2.5$.
Using equation (9) together with equations (2)--(4), we find that
the final mass fractions of $\alpha$-particles and neutrons have to be
\begin{mathletters}
\begin{eqnarray}
X_{\alpha,f}&\approx&{Y_{e,f}-\bar Z/A\over 1/2-\bar Z/A},\\
X_{n,f}&\approx&{1-\bar A/A\over 1-2\bar Z/A}(1-2Y_{e,f}),
\end{eqnarray}
\end{mathletters}
respectively, in order to produce $r$-process nuclei with mass number $A$.
Comparing equations (8a) and (8b) with equations (10a) and (10b),
we see that the fractional change in the neutron mass fraction during
the $\alpha$-process is less than that in the $\alpha$-particle mass
fraction for $Y_{e,f}\approx Y_{e,i}<\bar Z/\bar A$. Obviously,
because neutrons carry no charge and the mean charge per nucleon for
$\alpha$-particles exceeds $\bar Z/\bar A$ for the heavy seed nuclei,
the final composition favors the presence of neutrons for 
$Y_{e,f}<\bar Z/\bar A$. Accordingly, we will present the analytic
treatment of the $\alpha$-process for two different cases:
$Y_{e,i}<\bar Z/\bar A$ and $Y_{e,i}>\bar Z/\bar A$. In both cases,
we will assume $Y_{e,f}\approx Y_{e,i}$.

As the temperature declines from 
$T_9\approx 5$ to 2.5, $\alpha$-particles and neutrons 
are partially assembled 
into heavy seed nuclei. We can describe the $\alpha$-process in terms of
the time evolution of the $\alpha$-particle and neutron abundances.
During the $\alpha$-process, the burning of $\alpha$-particles
mainly proceeds 
via the reaction sequence 
$^4{\rm He}(\alpha n,\gamma)^9{\rm Be}(\alpha,n)^{12}{\rm C}$.
The production of seed nuclei occurs through the efficient $\alpha$-capture
reactions starting with $^9{\rm Be}(\alpha,n)^{12}{\rm C}$. Consequently,
the rates of change in the $\alpha$-particle and neutron number fractions 
can be approximately written as
\begin{mathletters}
\begin{eqnarray}
\drvf{Y_\alpha}{t}&\approx&-FY_\alpha Y_9\rho 
N_A\langle\sigma v\rangle_{\alpha n},\\
\drvf{Y_n}{t}&\approx&-GY_\alpha Y_9\rho
N_A\langle\sigma v\rangle_{\alpha n},
\end{eqnarray}
\end{mathletters}
respectively, where $Y_9$ is the number fraction of $^9$Be, and
$N_A\langle\sigma v\rangle_{\alpha n}$ is the reaction rate for 
$^9$Be($\alpha,n$)$^{12}$C in units of cm$^3$ s$^{-1}$ g$^{-1}$.
In equations (11a) and (11b), 
$F$ and $G$ are the numbers of $\alpha$-particles and neutrons that 
make up a typical heavy seed nucleus. For a seed distribution with
mean proton number $\bar Z$ and mean mass number $\bar A$, we have
$F\approx\bar Z/2$ and $G\approx\bar A-2\bar Z$.   

Due to its low $Q$-value of only 1.573 MeV, the reaction 
$^4{\rm He}(\alpha n,\gamma)^9$Be is tightly balanced by its reverse reaction
essentially over the entire temperature range $5\gtrsim T_9\gtrsim 2.5$.
According to statistical equilibrium, the number fraction of $^9$Be during
the $\alpha$-process is given by 
\begin{equation}
Y_9\approx 8.66\times 10^{-11}
Y_\alpha^2 Y_n \rho_5^2T_9^{-3} \exp(18.26/T_9).
\end{equation}
Because the density $\rho$ is related to the temperature $T_9$ through
the constant entropy $S$, and $N_A\langle\sigma v\rangle_{\alpha n}$ depends
on $T_9$ only, equations (11a) and (11b) can be expressed in a more
convenient form if we regard $Y_\alpha$ and $Y_n$ as functions of
temperature. Using equation (5), we have
\begin{mathletters}
\begin{eqnarray}
{dY_\alpha\over dT_9}&\approx&FY_\alpha^3Y_ng(T_9)\tau_{\rm dyn},\\
{dY_n\over dT_9}&\approx&GY_\alpha^3Y_ng(T_9)\tau_{\rm dyn},
\end{eqnarray}
\end{mathletters}
where $g(T_9)$ has the unit of s$^{-1}$, and is given by
\begin{equation}
g(T_9)\approx 8.66\times 10^{-6}\rho_5^3T_9^{-4}\exp(18.26/T_9)
N_A\langle\sigma v\rangle_{\alpha n}.
\end{equation} 

Now we can solve equations (13a) and (13b) for the two different cases
mentioned previously. For $Y_{e,i}<\bar Z/\bar A$, the final composition
favors the presence of neutrons. So we can approximately take
$Y_n\approx Y_{n,0}=X_{n,0}$ during the $\alpha$-process. In this case,
it is straightforward to solve equation (13a) and obtain
\begin{equation}
Y_{\alpha,f}^{-2}-Y_{\alpha,0}^{-2}\approx 2FY_{n,0}\tau_{\rm dyn}
\int_{2.5}^5g(T_9)dT_9.
\end{equation}
Likewise, for $Y_{e,i}>\bar Z/\bar A$, we can approximately take
$Y_\alpha\approx Y_{\alpha,0}=X_{\alpha,0}/4$ during the $\alpha$-process,
and solve equation (13b) to obtain
\begin{equation}
Y_{n,f}\approx Y_{n,0}\exp\left[-GY_{\alpha,0}^3\tau_{\rm dyn}
\int_{2.5}^5g(T_9)dT_9\right].
\end{equation}
Equations (15) and (16) implicitly constrain the combinations of
$Y_{e,i}$, $S$, and $\tau_{\rm dyn}$, for which the $\alpha$-process
can give a sufficient neutron-to-seed ratio for production of $r$-process
nuclei with mass number $A$.

To proceed further, we approximate the constant entropy during the
adiabatic expansion as $S\approx C(T_9)T_9^3/\rho_5$,
with $C(T_9)$ decreasing from 3.33 at
$T_9\gtrsim 5$ to 1.21 at $T_9\lesssim 1$. So equation (14) can be 
rewritten as
\begin{equation}
\label{negt9}
g(T_9)\approx 8.66\times 10^{-6}S^{-3}C(T_9)^3T_9^5
\exp(18.26/T_9)N_A\langle\sigma v\rangle_{\alpha n}.
\end{equation}
We plot $g(T_9)S^3$ as a function of $T_9$ in Figure 11, assuming
$C(T_9)\approx 3.33$ and using the fitting formula for 
$N_A\langle\sigma v\rangle_{\alpha n}$ given by 
Wrean, Brune, \& Kavanagh (1994). 
As we can see from this figure, $g(T_9)$ decreases
monotonically with temperature over $2.5\lesssim T_9\lesssim 5$. 
At $T_9\approx 4$, $g(T_9)$ has
already fallen below half its value at $T_9\approx 5$. Clearly, the
main contribution to the integral $\int_{2.5}^5g(T_9)dT_9$ comes from
$4\lesssim T_9\lesssim 5$. This remains true even when the exact form
of $C(T_9)$ is used. For our analytic estimates, we can approximately
evaluate the integral with $C(T_9)\approx 3.33$, and obtain
\begin{equation}
\label{negt9i}
\int_{2.5}^5g(T_9)dT_9\approx 6.4\times 10^8S^{-3}\ {\rm s}^{-1}.
\end{equation}
Using equation (\ref{negt9i}) and assuming $Y_{e,f}\approx Y_{e,i}$, we can
rearrange equations (15) and (16) as
\begin{mathletters}
\begin{eqnarray}
S&\approx&\left\{{4\times 10^7\bar Z(1-2Y_{e,i})\over
\left[{{1/2-\bar Z/A}\over{Y_{e,i}-\bar Z/A}}\right]^2
-\left[{{1}\over{2Y_{e,i}}}\right]^2}\left({\tau_{\rm dyn}\over{\rm s}}
\right)\right\}^{1/3}, \ {\rm for} \ Y_{e,i} < {\bar Z\over \bar A},\\
S&\approx&Y_{e,i}\left\{{8\times 10^7(\bar A-2\bar Z)\over
\ln\left[{{1-2\bar Z/A}\over{1-\bar A/A}}\right]}\left({\tau_{\rm dyn}
\over{\rm s}}\right)\right\}^{1/3},\ {\rm for}\ Y_{e,i}>{\bar Z\over \bar A},
\end{eqnarray}
\end{mathletters}
\noindent{where we have also used equations (8a), (8b), (10a) and (10b).}

To compare our analytic results in equations (19a) and (19b) with
the numerical results for $A\approx 200$ in $\S3$, we take 
$\bar Z\approx 34$ and $\bar A\approx 90$ from the numerical survey,
and obtain
\begin{mathletters}
\begin{eqnarray}
\label{nesyet}
S&\approx&10^3\left\{{1-2Y_{e,i}\over\left[
{{0.33}\over{(Y_{e,i}-0.17)}}\right]^2-\left[{{1}\over{2Y_{e,i}}}\right]^2}
\left({t_{\rm exp}\over{\rm s}}\right)\right\}^{1/3}, \ {\rm 
for} \ Y_{e,i} < 0.38,\\ 
S&\approx&2\times 10^3Y_{e,i}\left({t_{\rm exp}
\over{\rm s}}\right)^{1/3}, \ {\rm for} \ Y_{e,i}>0.38,
\end{eqnarray}
\end{mathletters}
\noindent{where we have replaced $\tau_{\rm dyn}$ with $t_{\rm exp}\approx
1.28\tau_{\rm dyn}$. These analytic results are shown as open circles
connected by dotted lines in Figure 10. The general
agreement between our analytic results and the numerical survey using
equation (7) (filled circles connected by solid lines) is quite good.
Because $Y_\alpha$ and $Y_n$ can decrease during the $\alpha$-process
and effective burning of $\alpha$-particles may start at $T_9<5$,
we always tend to overestimate the entropy by holding either $Y_\alpha$
or $Y_n$ constant and doing the integral in equation (\ref{negt9i}) over the
entire temperature range $2.5\lesssim T_9\lesssim 5$ in our analytic
treatment. The largest discrepancies occur at $Y_e\gtrsim 0.47$
where we overestimate the entropy by $\sim(15$--50)\%.
Using the specific values of $\bar Z$ and $\bar A$ found in the 
numerical survey would slightly improve the agreement. Further improvement
could be obtained by solving equations (11a) and (11b) together instead
of approximately solving each for a specific case. However, these
improvements would add little to our understanding of the physics 
determining the neutron-to-seed ratio. We also notice that the same level
of agreement with the numerical survey using an exact equation of state
(filled squares) can be achieved if we use
$C(T_9)\approx 3$ instead of 3.33 to account for
the annihilation of electron-positron pairs into photons at $T_9<5$.}

From our analytic treatment of the $\alpha$-process, we can clearly see
the individual roles of the initial electron fraction, entropy, and
dynamic time scale in determining the neutron-to-seed ratio. In addition
to specifying the overall availability of neutrons (cf. eq. [8b]), 
the initial electron
fraction serves to direct the path of nuclear flow during the 
$\alpha$-process. As our analytic treatment indicates, the comparison
of $Y_{e,i}$ with the ratio $\bar Z/\bar A$ for the typical seed 
distribution reflects whether $\alpha$-particles or neutrons are
mainly consumed during the $\alpha$-process. The influence of entropy
is manifested through the density dependences of the equilibrium
abundance of $^9$Be (cf. eq. [12]) and the rate for burning 
$\alpha$-particles (cf. eq. [11a]). Physically, a high entropy means
many photons per baryon in radiation-dominated conditions. A significant
fraction of these photons can be on 
the high-energy tail of the Bose-Einstein
distribution, and therefore can maintain a low $^9$Be abundance through
the photo-disintegration reactions. In turn, this limits the overall
efficiency of burning $\alpha$-particles, and hence the production
of seed nuclei. The dynamic time scale, or the expansion time,
specifies the duration of the $\alpha$-process (cf. eq. [6]). Obviously,
the expansion time also acts to limit the production of seed nuclei. 
In general, a lower initial electron fraction, a higher entropy, and
a shorter expansion time all give a larger neutron-to-seed ratio.
It is most interesting to notice that for a given $Y_{e,i}$, the
composition resulting from the $\alpha$-process essentially only depends
on the combination $S^3/t_{\rm exp}$ (cf. eqs. [20a] and [20b]).
Consequently, the same neutron-to-seed ratio can be achieved for the same
initial electron fraction with an expansion time 8 times shorter
if the entropy is reduced by a factor of 2.

\section{Conclusions}

For reasonable assumptions regarding neutrino luminosity, neutron star mass
and radius, 
and the time history for $Y_e$, the nucleosynthesis resulting from the
analytic model developed in Paper I for the neutrino-driven wind does not
resemble the solar $r$-process, although a number of interesting species
in the mass range $90 \leq A \leq 120$ are produced. This failure may be a
consequence of important (but unknown) physics neglected in Paper I, or
our results may reflect the true nucleosynthesis from typical 
core-collapse-driven
supernovae (however, see the discussion concerning the ejection of the
wind material in $\S3$). 
Extra (but, so far, artificial) energy input to the wind beyond
the injection radius (where $\dot M$ is determined) does give a successful
$r$-process. Possible sources of this energy were discussed in Paper I.

A numerical survey has delineated the necessary combinations of key
parameters --- $Y_e$, entropy, and expansion time --- 
needed to produce the third
$r$-process peak (i.e., the platinum peak). 
High entropy is not a unique requirement. A shorter expansion
time also serves to limit the number of heavy seed nuclei produced and thus
increase the neutron-to-seed ratio. A lower $Y_e$ also leads
to a larger neutron-to-seed ratio.
The sensitivity of the neutron-to-seed ratio to $Y_e$ diminishes 
as one proceeds to shorter expansion times,
as does the sensitivity to the entropy. 
Specific values of $Y_e$, entropy, and expansion time to produce
the third $r$-process peak are given in Table 5 and 
Figure 10.

Approximate analytic formulae (eqs. [20a] and [20b]) were derived that give
the requisite entropy needed to 
produce the heavy $r$-process nuclei as functions of
$Y_e$ and expansion time. 
These equations can be used to gauge whether other unstudied supernova
models or other astrophysical environments are appropriate sites for 
making $r$-process nuclei.

Given that the standard wind models, without artificial modification, produce
a set of abundances distinct from the $r$-process, one must be concerned
about the observational consequences. One possibility already mentioned is
that important physics has still been omitted from the simple wind model ---
e.g., neutrino flavor mixing, added energy input from shocks, rotation, or
magnetic fields, convection, etc.  --- and that the conditions required
for the $r$-process may still ultimately be achieved in common 
core-collapse-driven
supernovae. Perhaps material having a much lower $Y_e$ than calculated by
Wilson in Woosley et al. (1994) is ejected (\cite{bur95}). Another
possibility which must be seriously considered however, is that the
$r$-process has more than one important site and neutron exposure.

Sneden et al. (1996) and Cowan et al. (1996) have observed elements
attributed to the third $r$-process peak in the metal poor halo giant stars
HD 126238 ([Fe/H]$=-1.7$) and CS 22892-052 ([Fe/H]$=-3.1$). In HD 126238, the
scaled solar abundances of both Os and Pt have
been clearly observed. Both elements are made almost exclusively by the
$r$-process and fit the solar $r$-process abundance pattern. 
Similar results hold for the
more metal poor star CS 22892-052, although Pt was not observed and the
detection of Os is less certain. Coupled with previous data, the fit to the
solar $r$-process pattern for all elements between Ba and Os is striking,
suggesting that, for this range of nuclei, the solar $r$-process abundance
distribution appears to be made in its entirety in the progenitor(s) of these
metal poor stars. This result argues for a primary production scenario. Due to
the star's very low metallicity, especially so for CS 22892-052, the 
observed $r$-process abundance pattern
probably arose from only a few supernovae.

Interestingly, the abundance pattern for CS 22892-052 shows that elements in
the first neutron capture peak (Sr, Y, and Zr) are below their scaled solar
$r$-process fractions relative to Pt, Os, and Th, yet well above the
the iron group. Thus locally the $r$-process abundances seem solar,
but globally they are not. Sneden et al. (1996) suggest that the bulk of the
solar abundance of Sr, Y, and Zr is due to the  $s$-process, but these 
nuclei are also easily produced in the neutrino-driven wind (Figure 1
here; \cite{wh92}; \cite{wjt94}; \cite{wwmhm94}). This is consistent
with the existence of two sources for the $r$-process, one responsible for
production of the $r$-isotopes for $Z < 56$, including those in the $N=50$
peak, and the other for the heavier elements, possibly operating in another
site or a higher entropy version of the same site.

A very different type of $r$-process would arise also for much shorter
expansion times, as might occur in accretion induced collapse where the
wind is not slowed down by a massive overlying mantle (\cite{wobar92};
\cite{wh92}). A high entropy $r$-process with short expansion time
could occur through ejection by relativistic jets in coalescing neutron
stars (\cite{ruf96}). With short expansion times, the duration of
the neutron capture phase of the $r$-process may require
special consideration. If the material undergoing
nucleosynthesis cannot be slowed down during the neutron
capture phase, the $r$-process may have to be accelerated
by $\nu_e$ captures on heavy nuclei (Nadyozhin \& Panov 1993).
Ultimately, observational signatures of these
very distinct physical processes may be needed to resolve the nature and
site(s) of the $r$-process.

We would like to thank George Fuller and Gail McLaughlin for very
helpful discussions concerning the evolution of $Y_e$ in the
neutrino-driven wind and also for providing the
neutrino capture rates used in this work. We also gratefully acknowledge 
the Institute for Nuclear Theory at the University of Washington and
the Max-Planck Institute for Astrophysik for their 
generous hospitality during completion of this paper. 
This work was supported by NSF grant No. AST 94-17161 at UCSC. Woosley
was also supported by an Alexander von Humboldt Stiftung in Germany.
Y.-Z. Qian was supported by the D. W. Morrisroe Fellowship at Caltech.

% newr.apj.latex figures
% winds
\begin{figure}
%here is the overpro plot for winds1
%\epsfxsize=12.5 true cm
%\epsffile[0 0 612 612] 
%{winds1.eps}
\caption{Nucleosynthesis in
KEPLER wind model 10A: M $=1.4$ M$_\odot$, R $=10$ km,
$L_{\nu{\rm tot}}=1.8\times 10^{52}$ erg s$^{-1}$. Nuclei with
N = 50 closed neutron-shell dominate the nucleosynthesis.}\label{nfws1}
\end{figure}
%\clearpage

\begin{figure}
%here is the overpro plot for winds2
%\epsfxsize=12.5 true cm
%\epsffile[0 0 612 612] 
%{winds2.eps}
\caption{Nucleosynthesis in
KEPLER wind model 10B: M $=1.4$ M$_\odot$, R $=10$ km,
$L_{\nu{\rm tot}}=6.0\times 10^{51}$ erg s$^{-1}$.}\label{nfws2}
\end{figure}
%\clearpage

\begin{figure}
%here is the overpro plot for winds3
%\epsfxsize=12.5 true cm
%\epsffile[0 0 612 612] 
%{winds3.eps}
\caption{Nucleosynthesis in
KEPLER wind model 10C: M $=1.4$ M$_\odot$, R $=10$ km,
$L_{\nu{\rm tot}}=3.6\times 10^{51}$ erg s$^{-1}$.}\label{nfws3}
\end{figure}
%\clearpage

%windr

\begin{figure}
%here is the overpro plot for windr1
%\epsfxsize=12.5 true cm
%\epsffile[0 0 612 612] 
%{windr1.eps}
\caption{Nucleosynthesis in
KEPLER wind model 10D: M $=2.0$ M$_\odot$, R $=10$ km,
$L_{\nu{\rm tot}}=1.8\times 10^{52}$ erg s$^{-1}$.}\label{nfwr1}
\end{figure}
%\clearpage

\begin{figure}
%here is the overpro plot for windr2
%\epsfxsize=12.5 true cm
%\epsffile[0 0 612 612] 
%{windr2.eps}
\caption{Nucleosynthesis in
KEPLER wind model 10E: M $=2.0$ M$_\odot$, R $=10$ km,
$L_{\nu{\rm tot}}=6.0\times 10^{51}$ erg s$^{-1}$.}\label{nfwr2}
\end{figure}
%\clearpage

\begin{figure}
%here is the overpro plot for windr3
%\epsfxsize=12.5 true cm
%\epsffile[0 0 612 612] 
%{windr3.eps}
\caption{Nucleosynthesis in
KEPLER wind model 10F: M $=2.0$ M$_\odot$, R $=10$ km,
$L_{\nu{\rm tot}}=3.6\times 10^{51}$ erg s$^{-1}$.}\label{nfwr3}
\end{figure}
\clearpage

%windt
\begin{figure}
%here is the overpro plot for windt1
%\epsfxsize=12.5 true cm
%\epsffile[0 0 612 612] 
%{windt1.eps}
\caption{Nucleosynthesis in
KEPLER wind model 30A: M $=1.4$ M$_\odot$, 
R $=30$ km,
$L_{\nu{\rm tot}}=1.8\times 10^{53}$ erg s$^{-1}$.
These conditions lead to the production of species in the 
iron group.}\label{nfwt1}
\end{figure}
%\clearpage

%windt
\begin{figure}
%here is the overpro plot for windt2
%\epsfxsize=12.5 true cm
%\epsffile[0 0 612 612] 
%{windt2.eps}
\caption{Nucleosynthesis in
KEPLER wind model 30B: M $=1.4$ M$_\odot$, 
R $=30$ km,
$L_{\nu{\rm tot}}=6.0\times 10^{52}$ erg s$^{-1}$.}\label{nfwt2}
\end{figure}
%\clearpage

%windt
\begin{figure}
%here is the overpro plot for windt3
%\epsfxsize=12.5 true cm
%\epsffile[0 0 612 612] 
%{windt3.eps}
\caption{Nucleosynthesis in
KEPLER wind model 30C: M $=1.4$ M$_\odot$, 
R $=30$ km,
$L_{\nu{\rm tot}}=3.6\times 10^{52}$ erg s$^{-1}$.
The production of $^{64}$Zn (made as itself),
a unique signature of the neutrino-driven wind, 
suggests that such winds have likely occurred
in nature.}\label{nfwt3}
\end{figure}
%\clearpage

\begin{figure}
%here is the Y_e vs. entropy for the r-proc survey runs and the analytic 
%formula
%\epsfxsize=12.5 true cm
%\epsffile[0 0 612 612]
%{syet2.eps}
\caption{The combinations of $Y_e$, entropy, and expansion time 
required for the production of the $A\sim195$ $r$-process peak nuclei.
Circles connected by lines are for various fixed expansion 
times. Shown are the values derived 
in the numerical study using equation (7) (filled circles) 
and those from the analytic 
approximation (eqs. [20a] and [20b], open circles). 
The filled squares
represent results from the numerical survey that used an exact
adiabatic equation of state.}\label{nfyesnak2}
\end{figure}
%\clearpage

\begin{figure}
%here is the function g(t9) 
%\epsfxsize=12.5 true cm
%\epsffile[0 0 612 612] 
%{gt9.eps}
\caption{The temperature dependent function $g({\rm T}_9)$  
(eq. [17]). The main contribution to
the integral of $g({\rm T}_9)$ comes from T$_9\gtaprx 4$.}\label{nfgt9}
\end{figure}\clearpage

%newr.apj.latex - tables
% wind models 10 a-c production factors and final quantities

\begin{deluxetable}{r c c c c c } 
\label{ntk10ac}
\tablecaption{Nucleosynthesis in Wind Models 10A-C}
\tablehead{
\colhead{ }                & \colhead{no-$\nu$}             & 
\colhead{$\nu$(np$^{\rm a}$)}    & \colhead{$\nu$(np,$\alpha^{\rm b}$)} &
\colhead{$\nu$(np,$^AZ^{\rm c}$)}& \colhead{$\nu$(np,$\alpha$,$^AZ$)} 
}
\startdata
\cutinhead{Model 10A: $S\sim 74$, $Y_{e,i}=0.465$, 
$\tau_{\rm dyn}=0.024$ s} 
 $Y_{e,f}$        &  .465 &
 .470             &  .470 &
 .470             &  .470 \nl
 $\bar A$         &  85.1 &
 81.0             &  80.7 &
 80.3             &  80.7 \nl
 $X_\alpha$ &  .456 &
 .501       &  .506 &
 .508       &  .504 \nl
 $X_n$            &  4.4(-12)$^{\rm d}$ &
 3.0(-12)         &  3.4(-12) &
 3.4(-12)         &  3.3(-12)\nl
\cutinhead{Model 10B: $S\sim 87$, $Y_{e,i}=0.372$, 
$\tau_{\rm dyn}=0.066$ s}
 $Y_{e,f}$        &  .373 & 
 .393             &  .395 & 
 .396             &  .394 \nl
 $\bar A$         &   98.5 &
 96.8             &  96.6  & 
 96.6             &  96.8 \nl
 $X_\alpha$       &  .167  & 
 .200             &  .204  &
 .205             &  .201 \nl
 $X_n$            &  4.73(-2) & 
 1.42(-2)         &  1.17(-2) &
 1.07(-2)         &  1.28(-2) \nl
\cutinhead{Model 10C: $S\sim 94$, $Y_{e,i}=0.354$, 
$\tau_{\rm dyn}=0.11$ s}
 $Y_{e,f}$        &  .355 & 
 .386             &  .390 & 
 .391             &  .387\nl
 $\bar A$         &  99.8 & 
 97.7             &  97.3 & 
 97.3             &  97.7 \nl
 $X_\alpha$       &  .135     & 
 .173             &  .178     & 
 .180             &  .174 \nl
 $X_n$            &  7.75(-2) & 
 2.10(-2)         &  1.57(-2) & 
 1.42(-2)         &  1.91(-2)
\tablenotetext{a}{$\nu_e$ and $\bar\nu_e$ captures on nucleons (np)}
\tablenotetext{b}{neutral-current neutrino spallations
on $\alpha$-particles ($\alpha$)}
\tablenotetext{c}{$\nu_e$ captures on heavy nuclei ($^AZ$)}
\tablenotetext{d}{$x(y)\equiv x\times 10^y$}
\enddata
\end{deluxetable}
\clearpage

% r wind models 10 d-f

\begin{deluxetable}{r c c c c c } 
\label{ntk10df}
\tablecaption{Nucleosynthesis in Wind Models 10D-F} 
\tablehead{
\colhead{ }                & \colhead{no-$\nu$}             & 
\colhead{$\nu$(np$^{\rm a}$)}    & \colhead{$\nu$(np,$\alpha^{\rm b}$)} &
\colhead{$\nu$(np,$^AZ^{\rm c}$)}& \colhead{$\nu$(np,$\alpha$,$^AZ$)} 
}
\startdata
\cutinhead{Model 10D: $S\sim 109$, $Y_{e,i}=0.465$, 
$\tau_{\rm dyn}=0.024$ s}
 $Y_{e,f}$        &  .465  & 
 .470             &  .471  & 
 .471             &  .471 \nl
 $\bar A$         &  90.0  & 
 89.0             &  89.0  & 
 88.9             &  88.9  \nl
 $X_\alpha$       &  .642     & 
 .679             &  .682     &
 .683             &  .683  \nl
 $X_n$            &  1.78(-5)$^{\rm d}$ & 
 1.92(-6)         &  1.74(-6) & 
 1.22(-6)         &  1.22(-6) \nl 
\cutinhead{Model 10E: $S\sim 129$, $Y_{e,i}=0.372$, 
$\tau_{\rm dyn}=0.066$ s}
 $Y_{e,f}$        &  .372  & 
 .396             &  .398  & 
 .399             &  .399  \nl
 $\bar A$         &  98.0  & 
 97.2             &  97.1  & 
 97.2             &  97.2  \nl
 $X_\alpha$       &  .302 & 
 .345             &  .348 & 
 .348             &  .348 \nl
 $X_n$            &  9.54(-2) & 
 5.10(-2)         &  4.80(-2) & 
 4.66(-2)         &  4.66(-2)\nl
\cutinhead{Model 10F: $S\sim 140$, $Y_{e,i}=0.354$, 
$\tau_{\rm dyn}=0.11$ s}
 $Y_{e,f}$         &  .355  & 
 .390              &  .393  & 
 .393              &  .393  \nl
 $\bar A$          &  99.1  & 
 98.1              &  97.9  & 
 98.0              &  98.0  \nl
 $X_\alpha$        &  .262  & 
 .315              &  .320  & 
 .321              &  .321  \nl
 $X_n$             &  1.26(-2)  & 
 5.69(-2)          &  5.16(-2)  & 
 4.99(-2)          &  4.99(-2)  
\tablenotetext{a}{$\nu_e$ and $\bar\nu_e$ captures on nucleons (np)}
\tablenotetext{b}{neutral-current neutrino spallations 
on $\alpha$-particles ($\alpha$)}
\tablenotetext{c}{$\nu_e$ captures on heavy nuclei ($^AZ$)}
\tablenotetext{d}{$x(y)\equiv x\times 10^y$}
\enddata
\end{deluxetable}
\clearpage

% t wind models 30 a-c

\begin{deluxetable}{r c c c c c } 
\label{ntk30ac}
\tablecaption{Nucleosynthesis in Wind Models 30A-C}
\tablehead{
\colhead{ }                & \colhead{no-$\nu$}             & 
\colhead{$\nu$(np$^{\rm a}$)}    & \colhead{$\nu$(np,$\alpha^{\rm b}$)} &
\colhead{$\nu$(np,$^AZ^{\rm c}$)}& \colhead{$\nu$(np,$\alpha$,$^AZ$)} 
}
\startdata
\cutinhead{Model 30A: $S\sim 24$, $Y_{e,i}=0.510$, $\tau_{\rm dyn}=0.032$ s}
 $Y_{e,f}$         &  .510  & 
 .509              &  .509  & 
 .508              &  .508  \nl
 $\bar A$          &  56.1  & 
 56.1              &  56.1  & 
 56.1              &  56.1  \nl
 $X_\alpha$        &  .271  & 
 .271              &  .271  & 
 .272              &  .272  \nl
 $X_n$             &  3.(-18)$^{\rm d}$  & 
 1.(-15)           &  1.(-15)  &
 1.(-15)           &  1.(-15)  \nl
\cutinhead{Model 30B: $S\sim 26$, $Y_{e,i}=0.488$, $\tau_{\rm dyn}=0.075$ s}
 $Y_{e,f}$         &  .488  & 
 .489              &  .489  & 
 .490              &  .489  \nl
 $\bar A$          &  58.1  & 
 58.0              &  57.9  & 
 57.9              &  58.0  \nl
 $X_\alpha$        &  .220  & 
 .223              &  .224  & 
 .225              &  .223  \nl
 $X_n$             &  3.(-15)  & 
 2.(-15)           &  2.(-15)  &
 3.(-15)           &  3.(-15)  \nl
\cutinhead{Model 30C: $S\sim 28$, $Y_{e,i}=0.481$, $\tau_{\rm dyn}=0.12$ s}
 $Y_{e,f}$         &  .481  & 
 .483              &  .484  & 
 .484              &  .483  \nl
 $\bar A$          &  59.6  & 
 59.1              &  59.0  & 
 58.9              &  59.1  \nl
 $X_\alpha$        &  .181  & 
 .190              &  .194  & 
 .194              &  .191  \nl
 $X_n$             &  7.(-15)  & 
 7.(-15)           &  7.(-15)  &
 9.(-15)           &  1.(-14)  
\tablenotetext{a}{$\nu_e$ and $\bar\nu_e$ captures on nucleons (np)}
\tablenotetext{b}{neutral-current neutrino spallations
on $\alpha$-particles ($\alpha$)}
\tablenotetext{c}{$\nu_e$ captures on heavy nuclei ($^AZ$)}
\tablenotetext{d}{$x(y)\equiv x\times10^y$}
\enddata
\end{deluxetable}
\clearpage

\begin{deluxetable}{r c c c c c } 
\label{ntkx10fx}
\tablecaption{Nucleosynthesis in Modified Wind 
Model 10F with Extra Energy Input}
\tablehead{
\colhead{ }                & \colhead{no-$\nu$}             & 
\colhead{$\nu$(np$^{\rm a}$)}    & \colhead{$\nu$(np,$\alpha^{\rm b}$)} &
\colhead{$\nu$(np,$^AZ^{\rm c}$)}& \colhead{$\nu$(np,$\alpha$,$^AZ$)} 
}
\startdata
\cutinhead{Modified Model 10F: 
$S\sim 192$, $Y_{e,i}=0.354$, $\tau_{\rm dyn}=0.022$ s}
 $Y_{e,f}$         &  .354  & 
 .363              &  .363  & 
 .363              &  .363  \nl
 $\bar A$          &  89.7  & 
 91.0              &  91.0  & 
 91.0              &  91.0  \nl
 $X_\alpha$        &  .615  & 
 .629              &  .629  & 
 .629              &  .629  \nl
 $X_n$             &  .258  & 
 .240              &  .240  & 
 .240              &   .240  
\tablenotetext{a}{$\nu_e$ and $\bar\nu_e$ captures on nucleons (np)}
\tablenotetext{b}{neutral-current neutrino spallations
on $\alpha$-particles ($\alpha$)}
\tablenotetext{c}{$\nu_e$ captures on heavy nuclei ($^AZ$)}
\enddata
\end{deluxetable}
\clearpage

\begin{deluxetable}{c c c c c c c c }
\label{nt3}
\tablecaption{Results from the Numerical Survey Using 
$S\approx3.33T_9^3/\rho_5$}
\tablehead{
\colhead{$t_{\rm exp}$(s)} & \colhead{$Y_{e,f}$}& 
\colhead{$S$}& \colhead{$X_{\alpha,f}$}&
\colhead{$X_{n,f}$}& \colhead{$\bar{Z}$}& \colhead{$\bar{A}$}& \colhead{$n/s$}
}
\startdata
0.005& 0.200&   28&  0.074&   0.462&  35.8&  101.8& 101 \nl
     & 0.250&   58&  0.239&   0.396&  34.7&   97.0& 105 \nl
     & 0.300&   81&  0.391&   0.320&  33.9&   93.8& 104 \nl
     & 0.350&  103&  0.545&   0.242&  33.3&   91.3& 104 \nl
     & 0.400&  124&  0.696&   0.163&  32.9&   89.4& 103 \nl
     & 0.450&  145&  0.852&   0.083&  32.4&   87.2& 111 \nl
     & 0.460&  147&  0.882&   0.067&  32.2&   86.6& 112 \nl
     & 0.470&  144&  0.910&   0.050&  32.0&   85.8& 108 \nl
     & 0.480&  131&  0.942&   0.033&  32.2&   87.9& 116 \nl
     & 0.490&  117&  0.970&   0.016&  34.7&   97.1& 107 \nl
     & 0.495&  112&  0.983&   0.007&  41.1&  116.0&  85 \nl
0.025& 0.201&   50&  0.081&   0.466&  36.1&  101.8& 105 \nl
     & 0.251&   95&  0.232&   0.393&  35.2&   97.9& 103 \nl
     & 0.301&  136&  0.393&   0.320&  34.4&   94.7& 106 \nl
     & 0.350&  175&  0.551&   0.244&  33.7&   92.1& 110 \nl
     & 0.400&  210&  0.700&   0.164&  33.3&   90.0& 109 \nl
     & 0.450&  245&  0.852&   0.083&  32.9&   88.2& 111 \nl
     & 0.460&  250&  0.882&   0.066&  32.8&   87.8& 112 \nl
     & 0.470&  255&  0.912&   0.050&  32.7&   87.3& 115 \nl
     & 0.480&  249&  0.941&   0.033&  32.5&   86.5& 110 \nl
     & 0.490&  223&  0.970&   0.015&  35.8&   99.7& 100 \nl
     & 0.495&  210&  0.984&   0.007&  41.7&  117.2&  82 \nl
0.05 & 0.202&   55&  0.065&   0.457&  36.4&  102.6&  98 \nl
     & 0.252&  120&  0.236&   0.393&  35.4&   98.3& 104 \nl
     & 0.301&  170&  0.394&   0.320&  34.6&   95.1& 106 \nl
     & 0.351&  217&  0.550&   0.243&  34.0&   92.6& 108 \nl
     & 0.401&  263&  0.702&   0.164&  33.5&   90.4& 110 \nl
     & 0.451&  310&  0.852&   0.082&  33.2&   88.8& 111 \nl
     & 0.461&  315&  0.883&   0.066&  33.0&   88.2& 114 \nl
     & 0.471&  320&  0.912&   0.049&  33.0&   87.9& 113 \nl
     & 0.481&  323&  0.942&   0.033&  32.9&   87.4& 113 \nl
     & 0.490&  295&  0.970&   0.015&  35.4&   97.6& 101 \nl
     & 0.495&  278&  0.984&   0.006&  41.9&  117.7&  83 \nl
0.1  & 0.204&   70&  0.067&   0.453&  36.6&  103.2&  97 \nl
     & 0.253&  145&  0.228&   0.386&  35.7&   99.2&  99 \nl
     & 0.302&  210&  0.392&   0.316&  34.9&   95.8& 103 \nl
     & 0.352&  268&  0.548&   0.240&  34.2&   93.2& 106 \nl
     & 0.402&  325&  0.702&   0.162&  33.7&   90.9& 108 \nl
     & 0.451&  385&  0.853&   0.081&  33.4&   89.2& 109 \nl
     & 0.461&  390&  0.883&   0.064&  33.3&   88.8& 108 \nl
     & 0.471&  400&  0.913&   0.048&  33.2&   88.4& 110 \nl
     & 0.481&  410&  0.944&   0.032&  33.1&   87.9& 113 \nl
     & 0.491&  392&  0.973&   0.015&  32.5&   85.9& 108 \nl
     & 0.496&  369&  0.986&   0.006&  42.5&  118.8&  78 \nl
0.25 & 0.210&  105&  0.082&   0.443&  37.1&  104.5&  97 \nl
     & 0.258&  200&  0.239&   0.378&  36.2&  100.2&  99 \nl
     & 0.306&  285&  0.401&   0.309&  35.3&   96.8& 103 \nl
     & 0.355&  360&  0.555&   0.234&  34.6&   94.1& 104 \nl
     & 0.404&  435&  0.707&   0.157&  34.1&   91.6& 106 \nl
     & 0.453&  520&  0.859&   0.078&  33.7&   89.7& 110 \nl
     & 0.463&  530&  0.887&   0.061&  33.7&   89.6& 105 \nl
     & 0.473&  544&  0.919&   0.045&  33.5&   88.9& 111 \nl
     & 0.483&  565&  0.949&   0.029&  33.4&   88.4& 115 \nl
     & 0.493&  567&  0.978&   0.012&  33.1&   86.7& 107 \nl
     & 0.498&  532&  0.990&   0.004&  43.7&  121.8&  80 \nl
\enddata
\end{deluxetable}
\clearpage


\begin{thebibliography}{}

\bibitem[Anders \& Grevesse 1989]{ag89} Anders, E., \& Grevesse,
N. 1989, Geochim. Cosmochim. Acta, 53, 197

\bibitem[Burrows, Hayes, \& Fryxell 1995]{bur95}
Burrows, A., Hayes, J., \& Fryxell, B. A. 1995, \apj, 450, 830

\bibitem[Cowan, Thielemann, \& Truran 1991]{ctt91}
Cowan, J. J., Thielemann, F.-K., \& Truran, J. W. 1991, Phys. Rep., 208,
267

\bibitem[Cowan et al. 1996]{cow96}
Cowan, J. J., Sneden, C., Truran, J. W., \& Burris, D. L. 1996, \apjl,
in press.

\bibitem[Duncan, Shapiro, \& Wasserman 1986]{dsw86}
Duncan, R. C., Shapiro, S. L., \& Wasserman, I. 1986, \apj, 309, 141

\bibitem[Freiburghaus et al. 1996]{fri96}
Freiburghaus, Ch., Kolbe, E., Rauscher, T., Thielemann, F.-K.,
Kratz, K.-L., \& Pfeiffer, B. 1996, in {\it 
Fourth International Symposium on Nuclear Astrophysics}
University of Notre Dame, USA

\bibitem[Fuller \& Meyer 1995]{fulmey95} 
Fuller, G. M., \& Meyer, B. S. 1995, \apj, 453, 792

\bibitem[Hoffman et al. 1996]{hwfm96}
Hoffman, R. D., Woosley, S. E., Fuller, G. M., \& Meyer, B. S. 1996, 
\apj, 460, 478

\bibitem[Howard et al. 1993]{how93}
Howard, W. M., Goriely, S., Rayet, M., \& Arnould, M. 1993, \apj, 417, 713

\bibitem[McLaughlin \& Fuller 1995]{macful95}
McLaughlin, G. C., \& Fuller, G. M. 1995, \apj, 455, 202

\bibitem[McLaughlin, Fuller, \& Wilson 1996]{mfw96}
McLaughlin, G. C., Fuller, G. M., \& Wilson, J. R. 1996, \apj, in press.

\bibitem[Mathews \&  Cowan 1990]{matcow90}
Mathews, G. J., \&  Cowan, J. J. 1990, {\it Nature}, 345, 491

\bibitem[Meyer 1995]{mey95}
Meyer, B. S. 1995, \apjl, 449, L55

\bibitem[Meyer \& Brown 1996]{mey96}
Meyer, B. S. \& Brown, J. 1996, in {\it 
Fourth International Symposium on Nuclear Astrophysics}
University of Notre Dame, USA

\bibitem[Nadyozhin \& Panov 1993]{nadpan93}
Nadyozhin, D. K. \& Panov, I. V. 1993, in Proc. Int. Symp.
on Weak and Electromagnetic Interactions in Nuclei (WEIN-92),
Ts. D. Vylov, ed. (World Scientific, Singapore, 1993) p. 479

\bibitem[Qian et al. 1993]{qian93}
Qian, Y.-Z., Fuller, G. M., Mathwes, G. J., Mayle, R. W., Wilson,
J. R., \& Woosley, S. E. 1993, Phys. Rev. Lett., 71, 1965

\bibitem[Qian \& Woosley 1996]{qw96}
Qian, Y.-Z., \& Woosley, S. E. 1996, (Paper I) \apj, in press.

\bibitem[Ruffert et al. 1996]{ruf96}
Ruffert, M., Janka, H.-Th., Takahashi, K., \& Sch\"afer, G. 1996, A\&A,
submitted.

\bibitem[Sneden et al. 1996]{sneden96}
Sneden, C., McWilliam, A., Preston, G. W., Cowan, J. J., Burris, D. L., 
and Amorsky, B. J. 1996, \apj, in press.

\bibitem[Takahashi, Witti, \& Janka 1994]{twj94}
Takahashi, K., Witti, J., \& Janka, H.-Th. 1994, A\&A, 286, 857

\bibitem[Takahashi 1996]{tak96}
Takahashi, K. 1996, in {\it Proceedings of the 8$^{th}$ Annual Ringburg 
Conference on Supernovae and Nucleosynthesis}.

\bibitem[Timmes, Woosley, \& Weaver 1995]{tww95}
Timmes, F. X., Woosley, S. E., \& Weaver, T. A. 1995, \apjs, 98, 617

\bibitem[Witti, Janka, \& Takahashi 1994]{wjt94}
Witti, J., Janka, H.-Th., \& Takahashi, K.  1994, A\&A, 286, 841

\bibitem[Woosley \& Baron 1992]{wobar92}
Woosley, S. E., \& Baron, E. 1992, \apj, 391, 228

\bibitem[Woosley \& Hoffman 1992]{wh92}
Woosley, S. E., \& Hoffman, R. D. 1992, \apj, 395, 202

\bibitem[Woosley \& Weaver 1993]{wowev93}
Woosley, S. E., \& Weaver, T. A. 1993, in {\sl Supernovae: Les
Houches Summer School Vol 54}, eds. S. Bludman, R. Mochkovitch, \& J.
Zinn-Justin, (Amsterdam: North Holland) 

\bibitem[Woosley \& Weaver 1995]{wowev95}
Woosley, S. E., \& Weaver, T. A. 1995, \apjs, 101, 181

\bibitem[Woosley et al. 1994]{wwmhm94}
Woosley, S. E., Wilson, J. R., Mathews, G. J., Hoffman, R. D., \& Meyer,
B. S. 1994, \apj, 433, 229 

\bibitem[Wrean, Brune, \& Kavanagh 1994]{wrean94}
Wrean, P. R., Brune, C. R. \& Kavanagh, R. W. 1994, Phys. Rev. C, 49, 1205

\end{thebibliography}
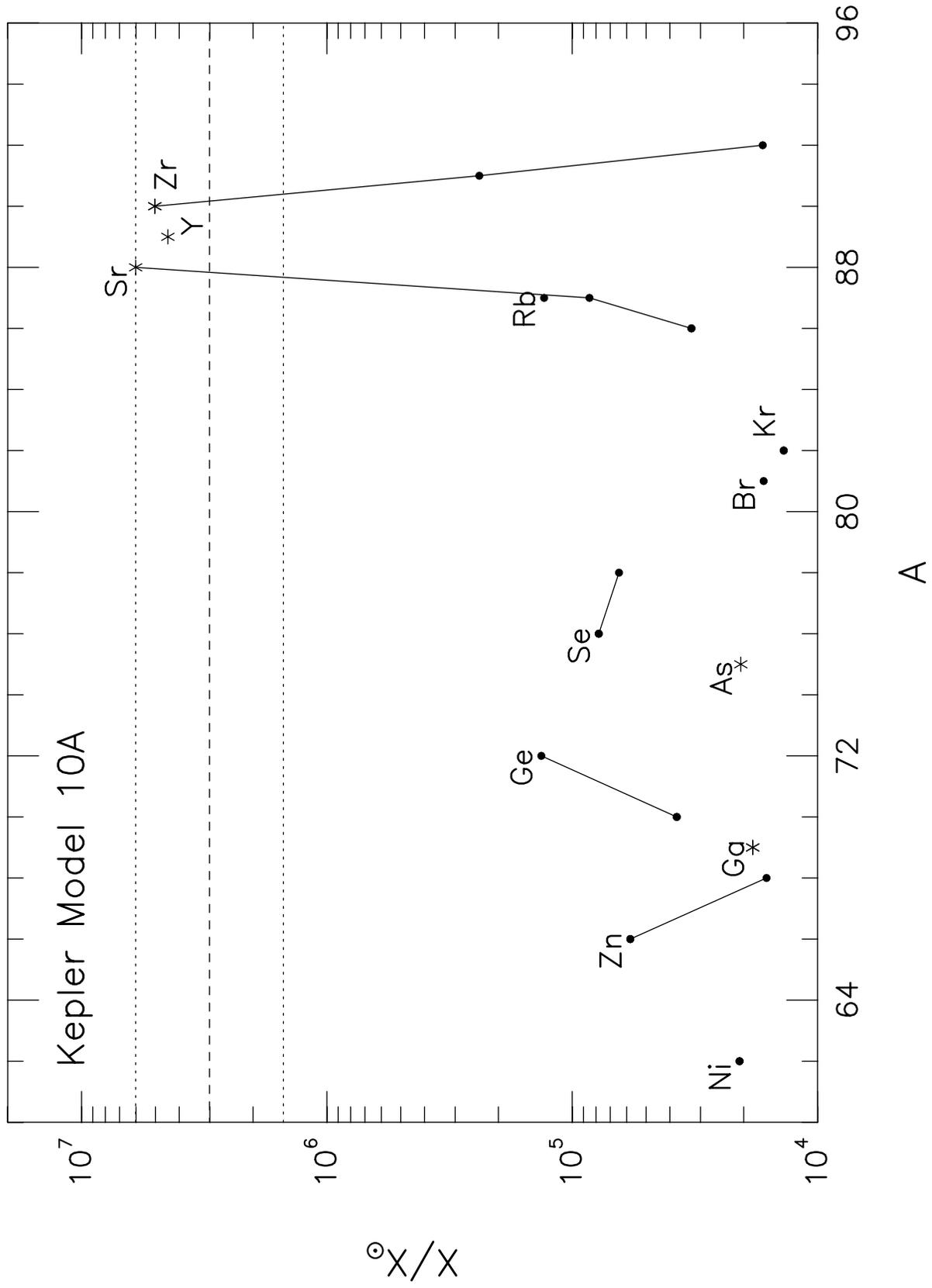
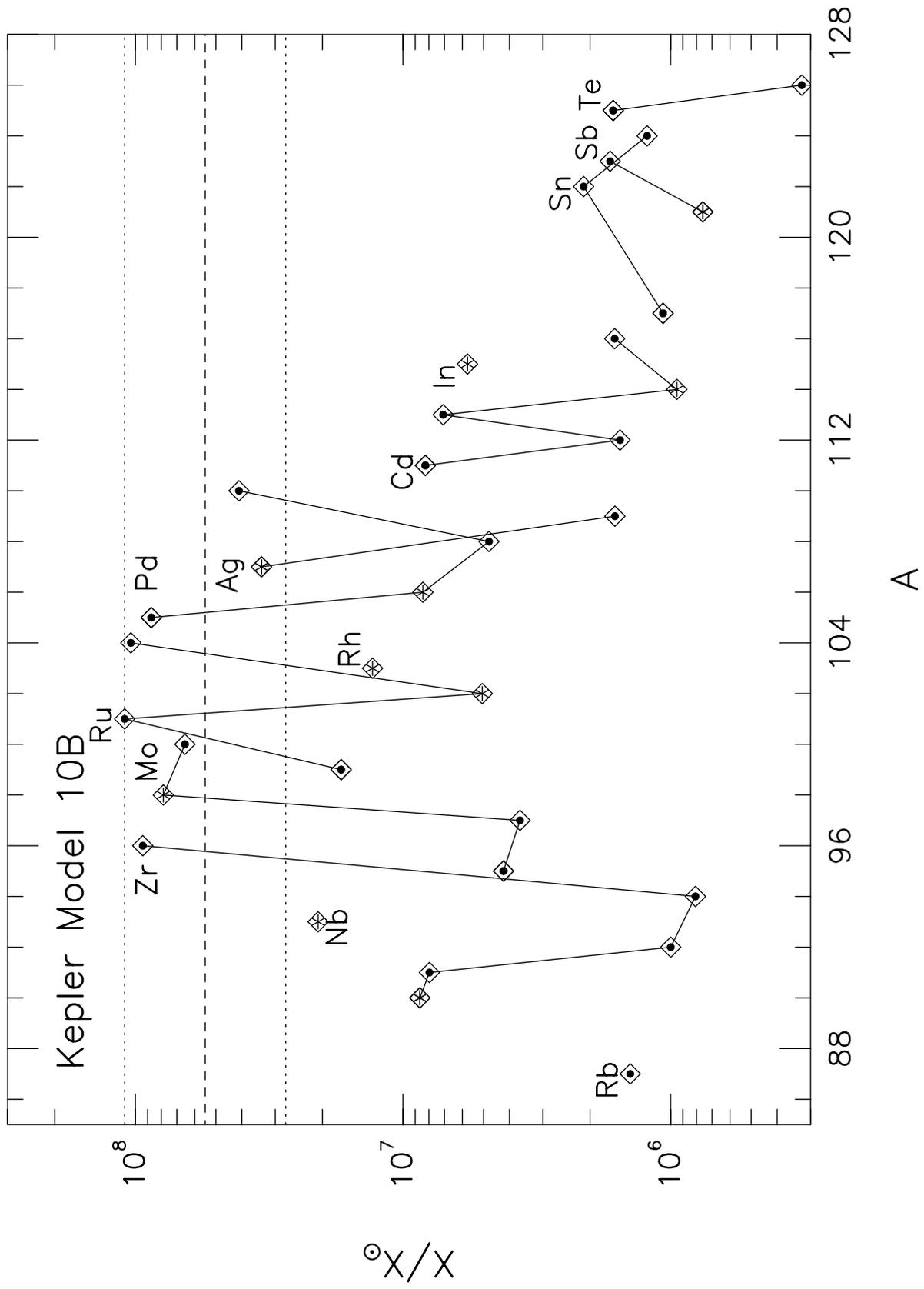
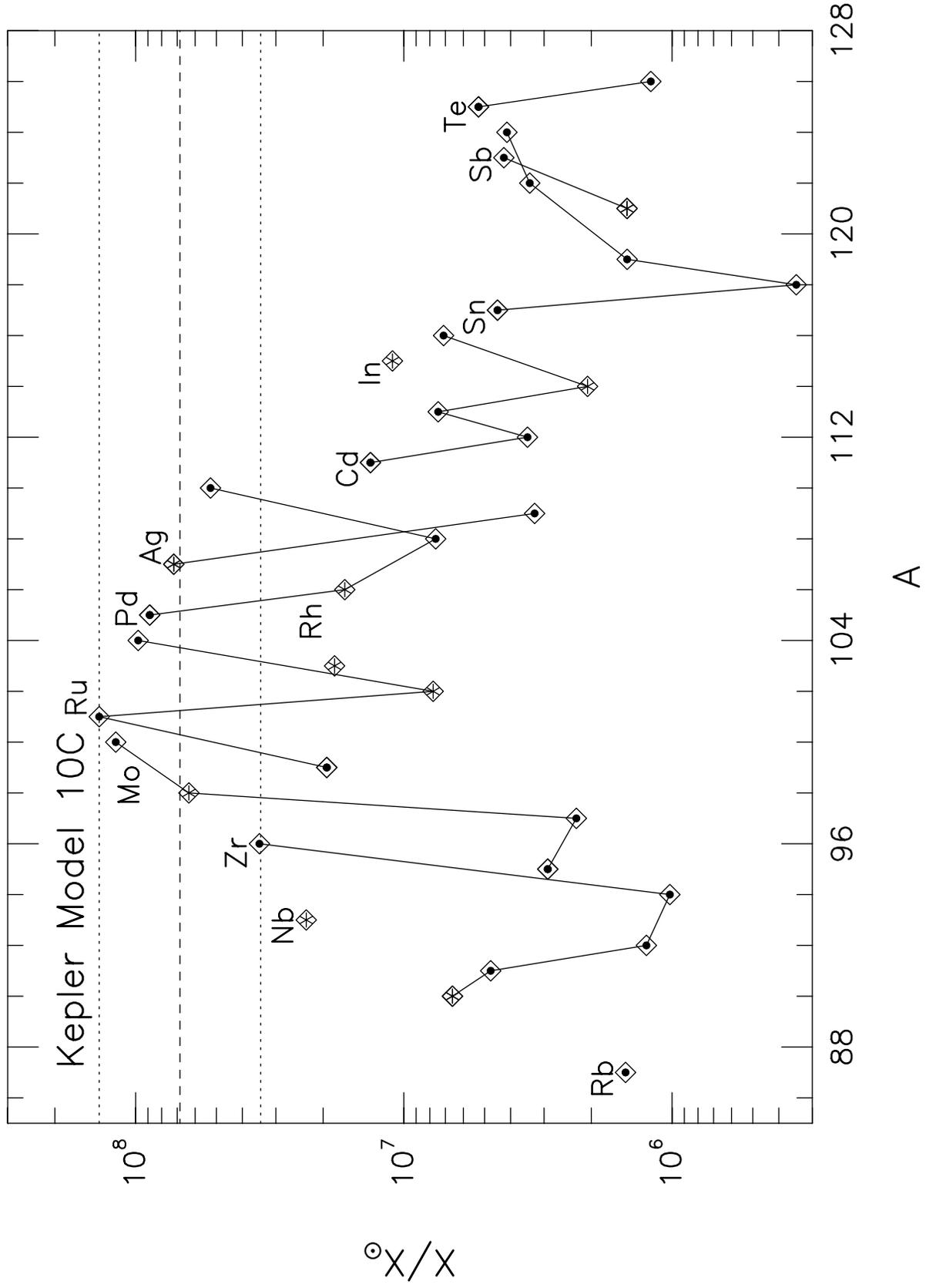
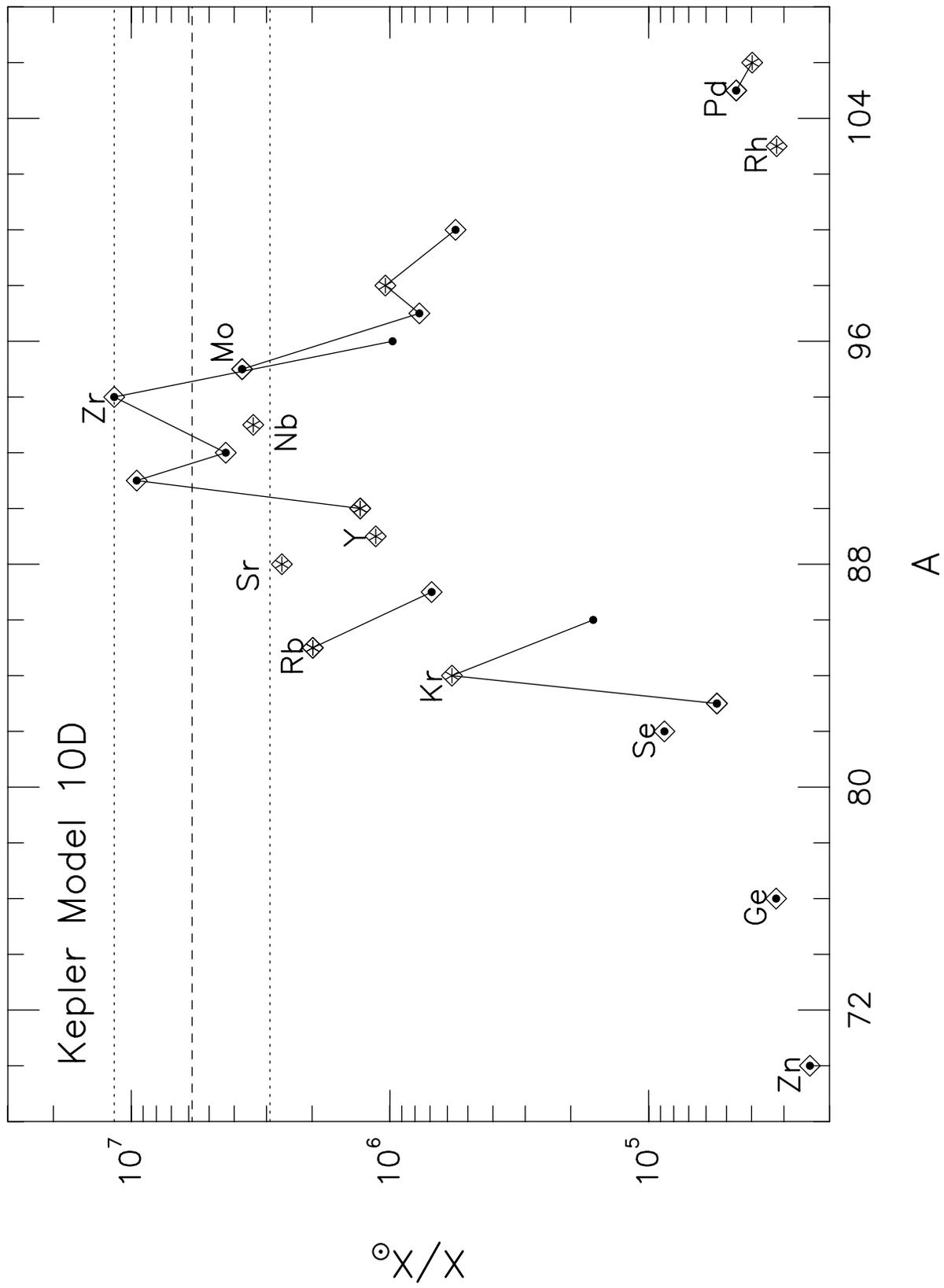
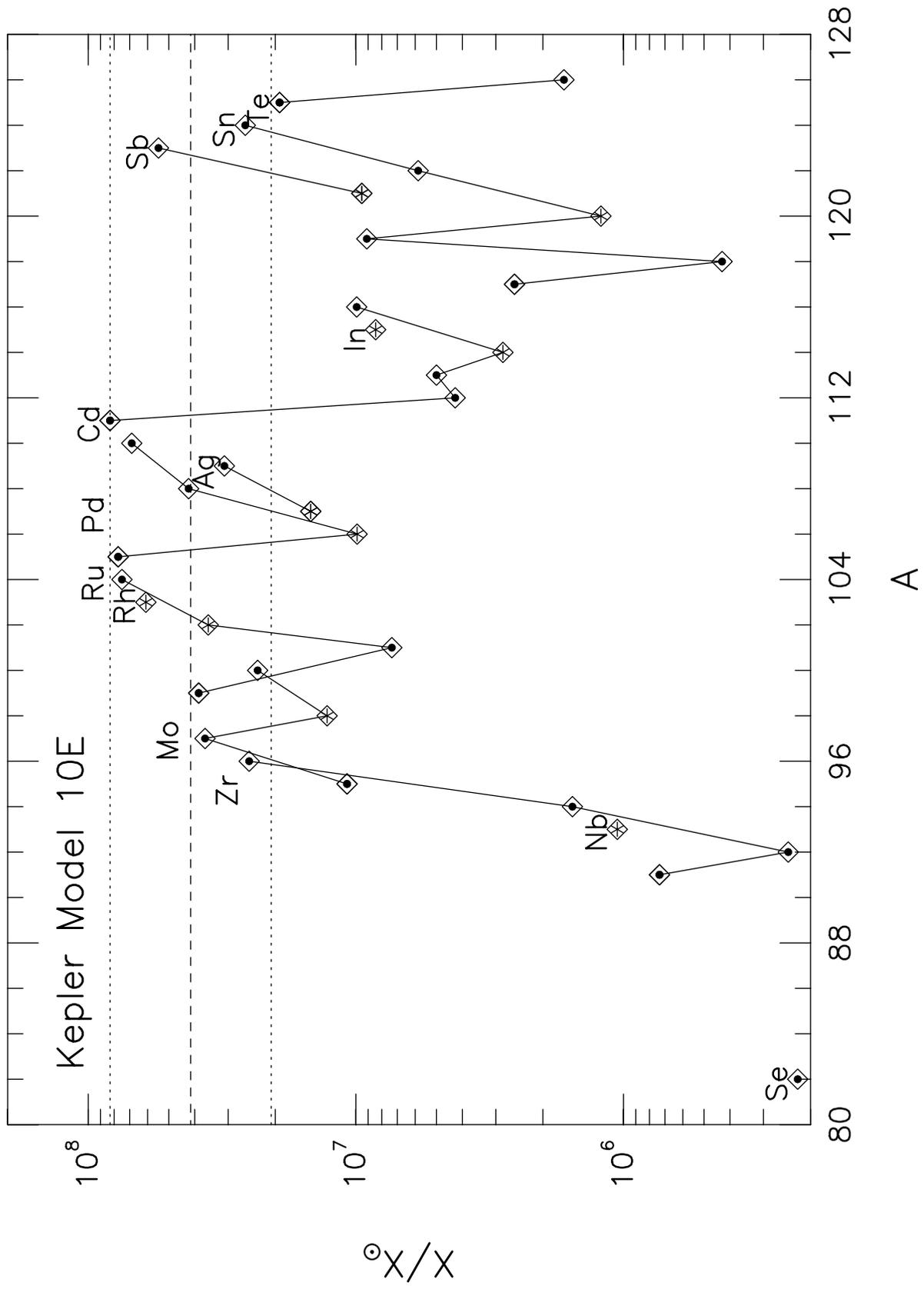
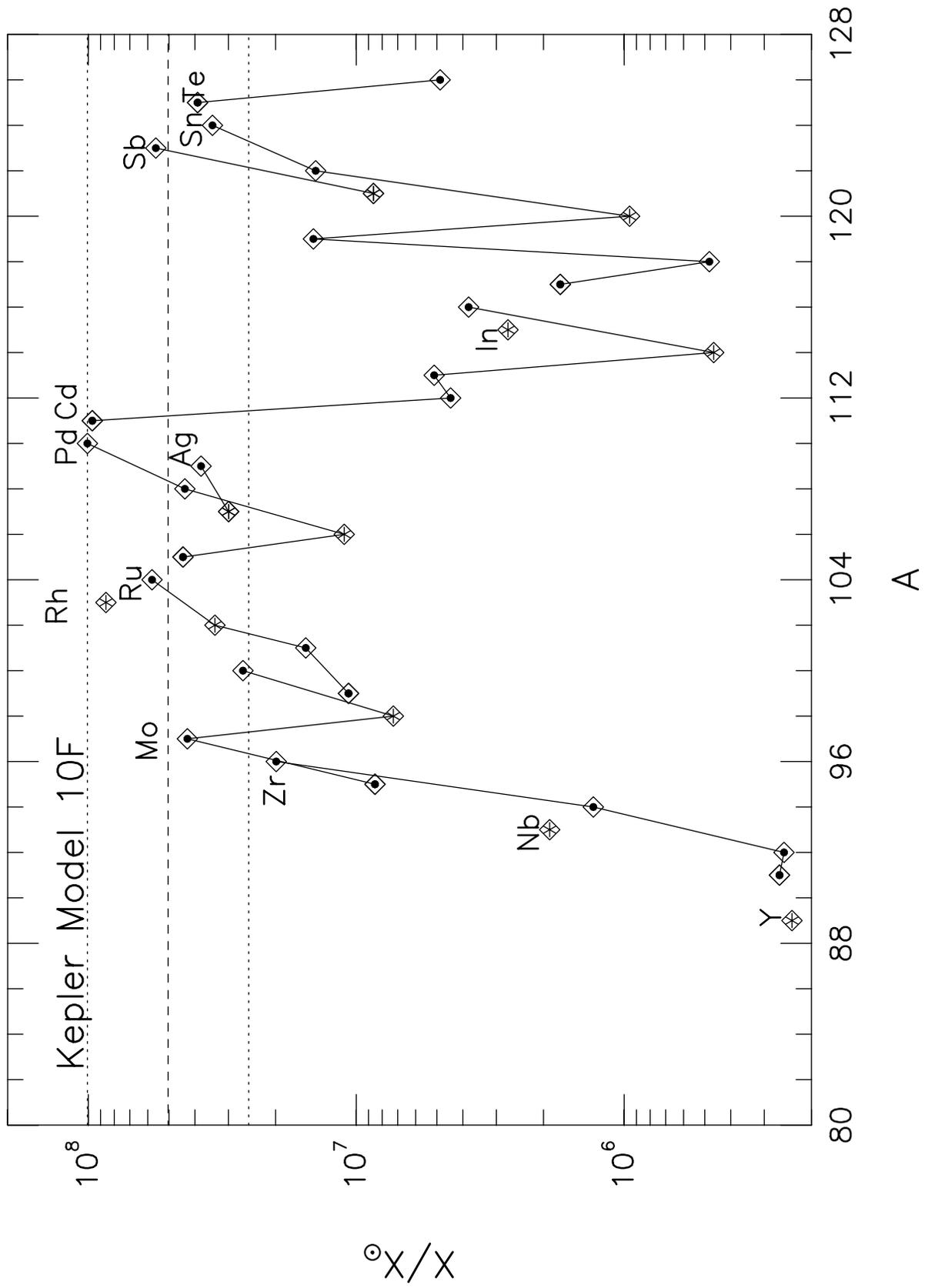
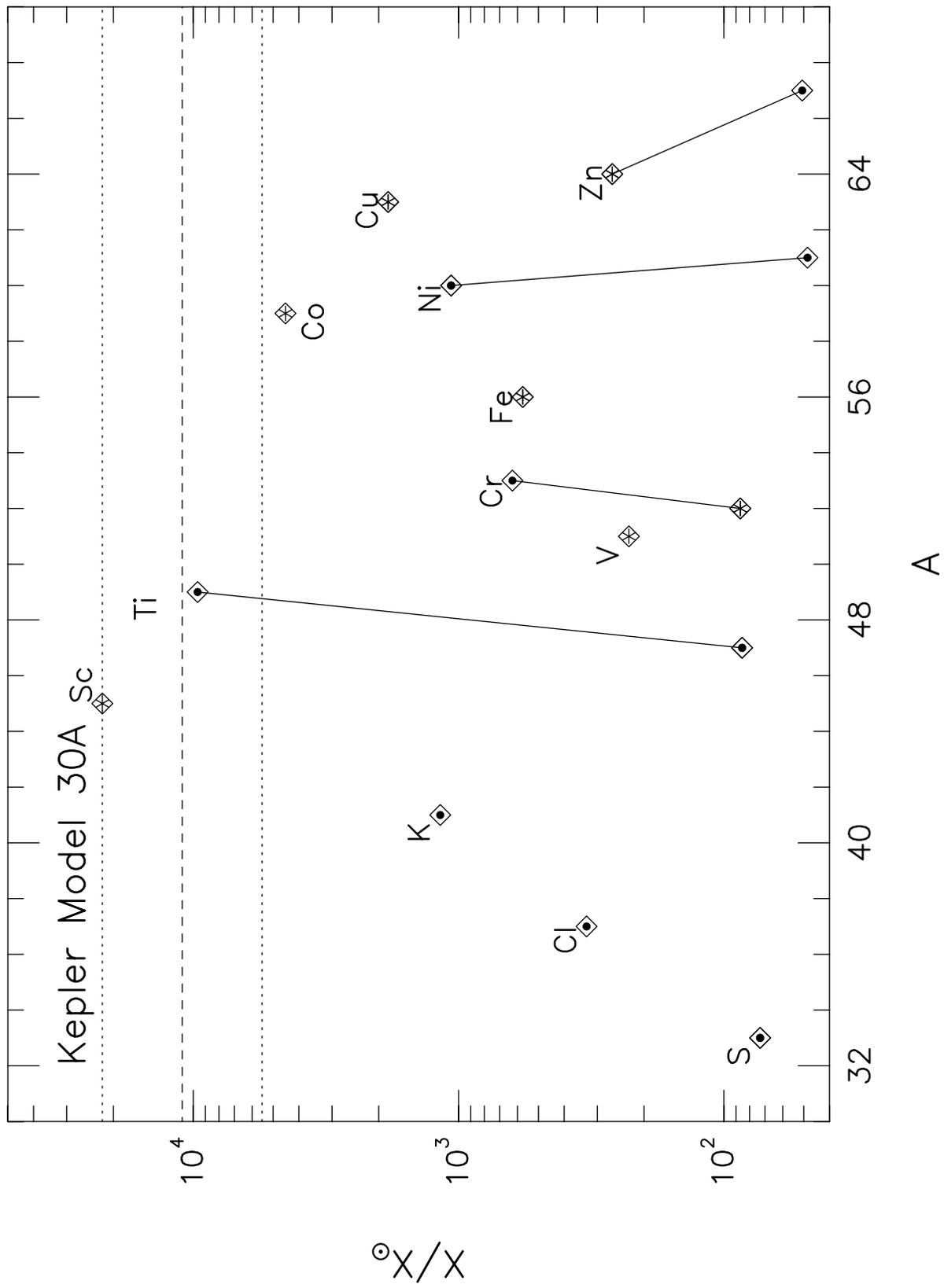
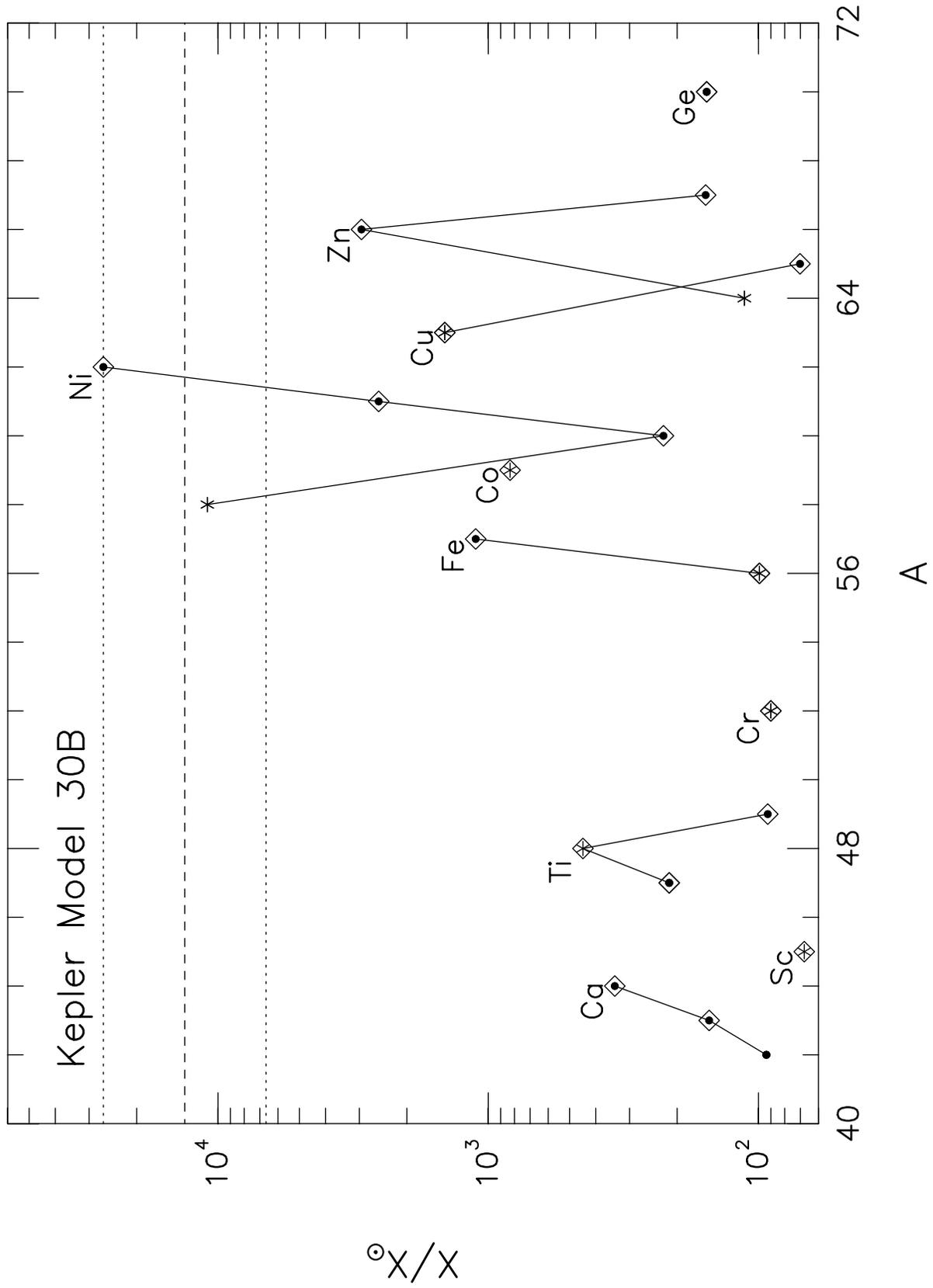
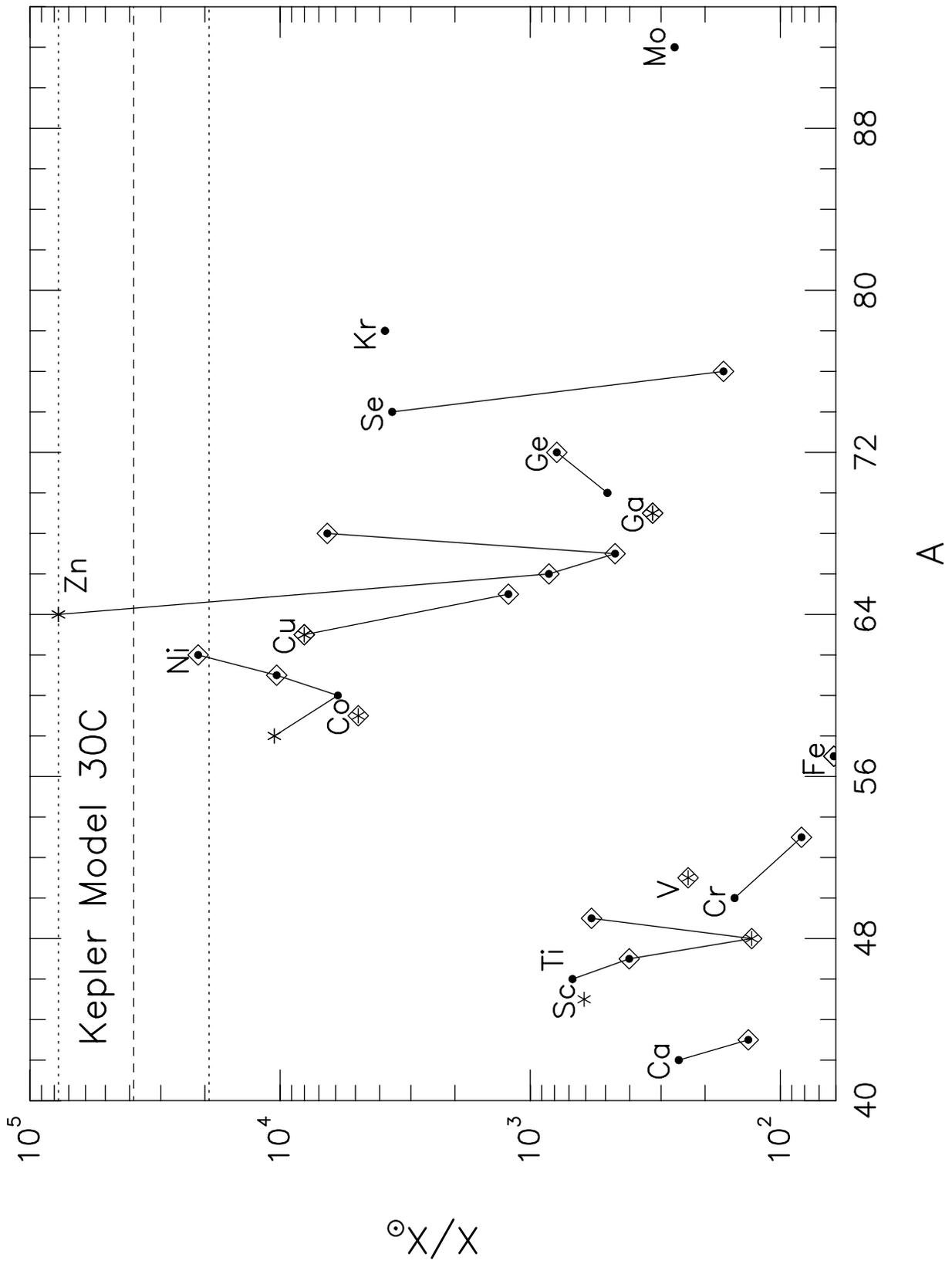
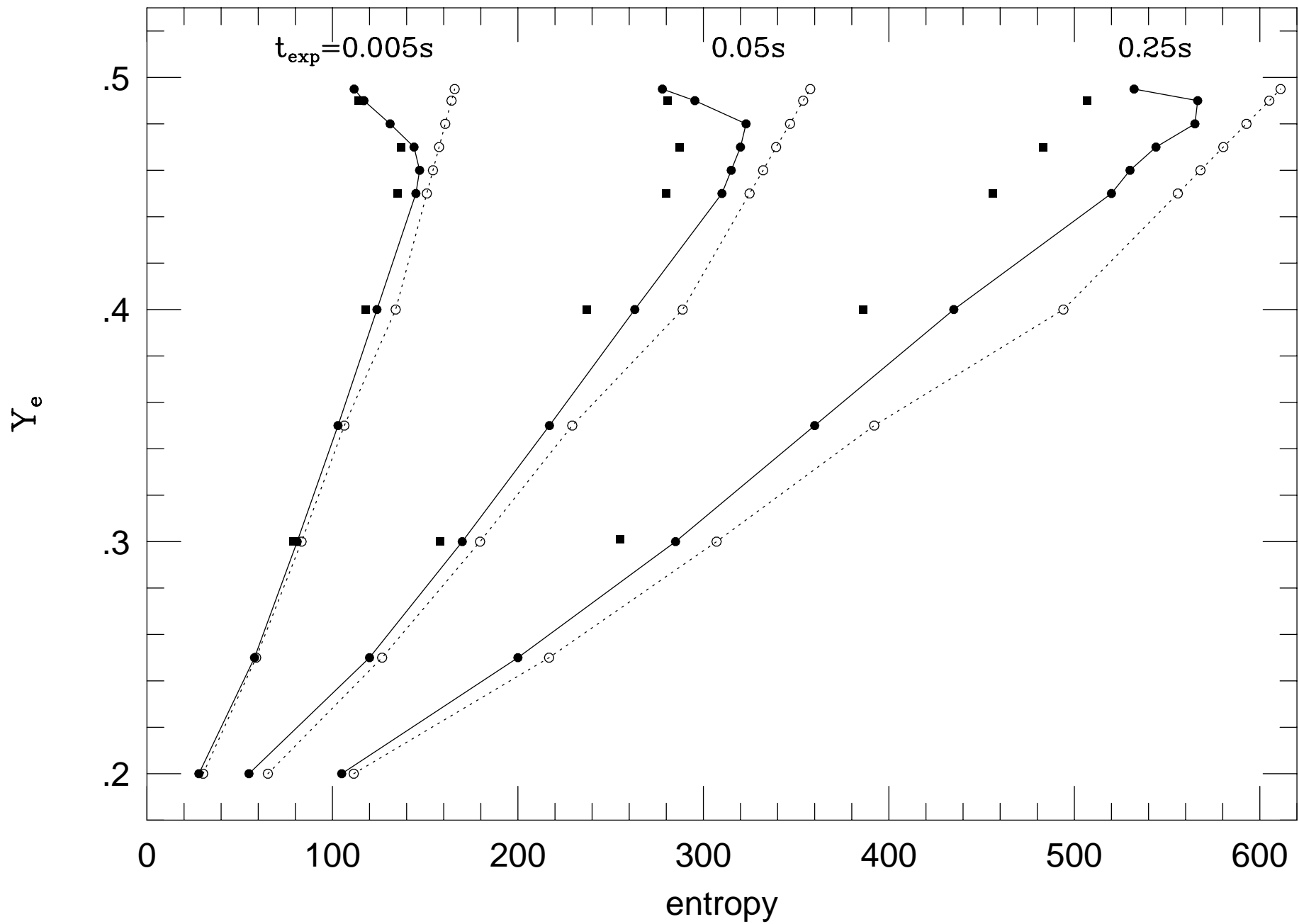
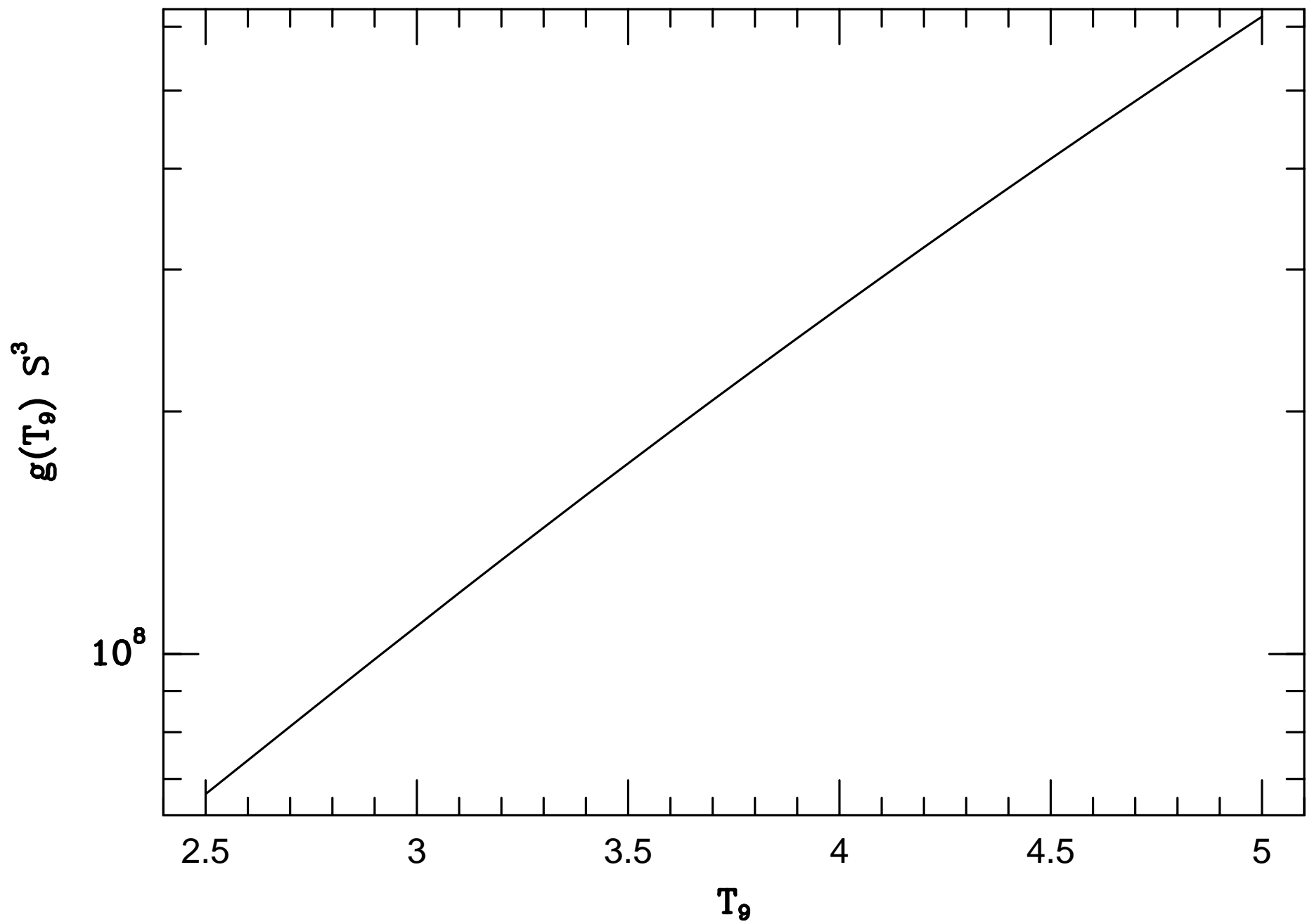
\end{document}